\documentclass[10pt]{article}
\usepackage{setspace}

\usepackage{amsmath}

\usepackage{graphicx}

\usepackage[numbers,sort&compress]{natbib}

\def\lsim{\mathrel{\rlap{\lower4pt\hbox{\hskip1pt$\sim$}}
    \raise1pt\hbox{$<$}}}                
\def\gsim{\mathrel{\rlap{\lower4pt\hbox{\hskip1pt$\sim$}}
    \raise1pt\hbox{$>$}}}                

\usepackage{epsfig}
\usepackage{color}

\title{\bf
Dynamical Analysis on Gene Activity 
in the Presence of Repressors and an
Interfering Promoter
}

\author{
{\bf Hiizu Nakanishi$^{1,2}$,
\bf Namiko Mitarai$^2$, and \bf Kim Sneppen}$^1$
\\\\ \it
{\rm $^1$} Niels Bohr Institute, Blegdamsvej 17, Dk 2100, Copenhagen, Denmark
 \\ \it
{\rm $^2$} Department of Physics, Kyushu University 33, Fukuoka 812-8582, Japan
}

\date{\today}

\begin{document}
\twocolumn
\maketitle

\abstract{ Transcription is regulated through interplay between
transcription factors, an RNA polymerase (RNAP), and a promoter.  Even
for a simple repressive transcription factor that disturbs promoter
activity at the initial binding of RNAP, its repression level is not
determined solely by the dissociation constant of transcription factor
but is sensitive to the time scales of processes in RNAP.
We first analyse the promoter activity under strong repression by a
slow binding repressor, in which case transcriptions occur in a burst,
followed by a long quiescent period while a repressor binds to the
operator; the number of transcriptions, the bursting and the
quiescent times are estimated by reaction rates.
We then examine interference effect from an opposing promoter, using the
correlation function of transcription initiations for a single promoter.
The interference is shown to de-repress the promoter because RNAP's from
the opposing promoter most likely encounter the repressor and remove it
in case of strong repression.
This de-repression mechanism should be especially prominent for the
promoters that facilitate fast formation of open complex with the
repressor whose binding rate is slower than $\sim$ 1/sec.  Finally, we
discuss possibility of this mechanism for high activity of promoter PR
in the hyp-mutant of lambda phage.

 }



\section*{Introduction}

The regulation of the activity of a particular gene involves a complex
interplay between a promoter, an RNA polymerase (RNAP), and one or
several transcription factors (TF) \cite{ptashne,roy}. Ignoring the
internal dynamics associated with transcription initiation, the
probability for obtaining a successful RNAP elongation initiation can
be estimated from an equilibrium unbinding ratio of
TF\cite{shea,sneppenbook}.
When internal steps in transcription initiations becomes sizeable
we need to consider the race between these steps and the
kinetics of TF binding.

The binding/unbinding rates of TF to bind to an operator is critically
influenced by competitive non-specific bindings \cite{winter,elf}.
Recent measurements of in vivo dynamics in an {\it E. coli} cell finds
that a single lac repressor needs between 60 and 360 seconds to locate
its operator \cite{elf}. For TF whose copy number is of the order of
10 to 100 per cell, a cleared operator can remain free for up to about
30 seconds.  In comparison, RNAP transcription initiation rates varies
considerably, and can be as fast as 1.8 transcription initiation per
second for a certain ribosomal promoter \cite{liang}.  Therefore,
there is ``room'' for effects associated to the race between first
bindings of a TF or an RNAP once the promoter is cleared.

In a number of both procaryotic and eucaryotic systems, the promoter
activity are not only influenced by TF, but are also modulated by
interfering promoters
\cite{ward,adhya,menendez,callen,greger,prescott,sneppen,dodd2007}.  For
example, the regulation between lytic and lysogenic maintenance
promoters in the P2 class of bacteriophages involves transcription
interferences (TI) as well as TF's that repress the promoter activities
\cite{callen}. And in lambdoid phages the initial lysis-lysogeny
decision is modulated by TI between the promoter PRE activated by CII
and the promoter PR repressed by CI.

Dodd {\it et al.}\cite{dodd2007} presented a framework to deal with TI
and multiple TF's, using an assumption about fast equilibrium
reactions of TF-binding and closed complex formation.  In the present
paper, we develop a formalism that deals with the competition between
time scales of TF binding/unbinding and transcription initiation
process, and examine the effect of interference.

\begin{figure}[tb]
\centerline{\epsfig{file=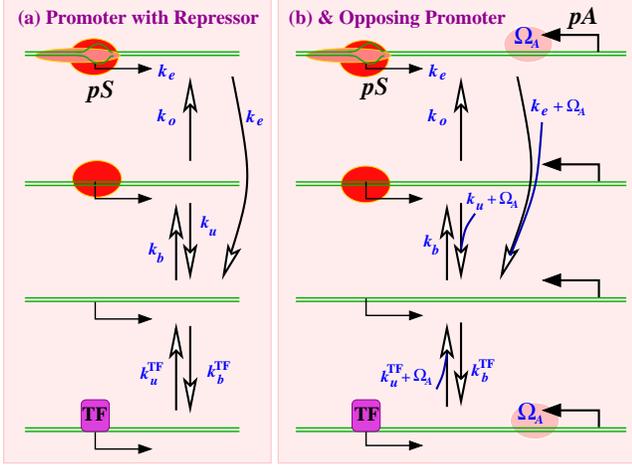,angle=0,width=8.5cm,clip=}}
\caption{\small (a) Model of promoter pS with a single TF that represses
the promoter by competitive binding to an operator that overlaps with
the promoter. The promoter activity is given in terms of the three-step
Hawley-McClure model for transcription initiation, with indicated
transition rates for formation of closed complex, that of open
complex, and elongation. (b) Same as in (a), but with addition of a
convergent promoter that interfere with both RNAP binding to pS and with
binding of TF.  \label{fig1}}
\end{figure}

Fig.\ref{fig1} shows a single promoter pS with an operator site for a
repressive TF (left panel), and with a convergent promoter pA (right
panel).  For both cases, we illustrate the three basic steps of
transcription initiation: (i) RNAP reversible binding to form a closed
complex, (ii) irreversible transition to open complex, and (iii)
initiation of transcription elongation.  The rates for these three steps
are promoter dependent \cite{hawley1982,buc,record}.  As for the initial
binding, given the fact that the maximum activity for ribosomal
promoters reaches 1.8 transcriptions per second\cite{liang}, the time
needed for an RNAP to diffuse to a promoter cannot be longer than $\sim
0.5$ sec.  Regarding the later steps where RNAP forms open complex and
subsequently initiates transcription to leave the promoter, their time
scales may vary a great deal from one promoter to
another\cite{hsu,knaus,carpousis,darzacq,lanzer}.

In the following, we will investigate in detail how these time scales play
together to determine the extent to which a promoter is sensitive to
repressors and to clearance due to the interference by elongating
RNAP's from other promoters\cite{applet}.


\section*{Models}

We study the promoter activity under influence of transcription
factor(TF) and transcription interference(TI) based on mathematical
analysis on simple models of promoter in the following three levels.
Our goal is to understand regulation of the three step model for
transcription initiation originally proposed by
Hawley-McClure\cite{hawley1982,buc}, but we also analyze its simplified
versions, i.e. the single step model and the two step model.  The
comparison of these three levels of models gives us intuitive
understanding of the promoter behavior.

\begin{figure}[t]
\centerline{\epsfig{file=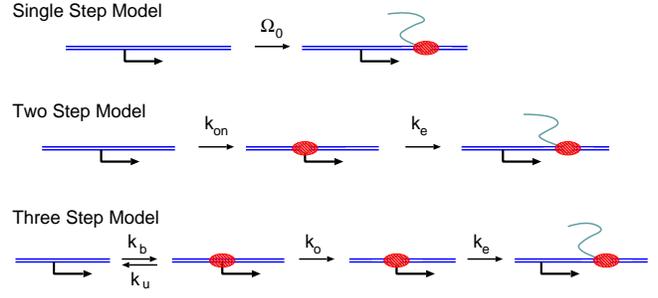,angle=0,width=8.5cm,clip=}}
\caption{\small Schematic illustrations for the single step model,
the two step model, and the three step model of the elongation initiation.}
\label{models}
\end{figure}

\subsection*{Three Models for Elongation Initiation}
Let us start by describing the bare models with neither TF nor TI
(Fig.\ref{models}).

i) {\em The single step model} of transcription initiation is the model
where the whole process is dominated by a slowest step, thus its
elongation initiation is represented by a simple Poissonian process with
the rate $\Omega_0$.  \vskip 1ex

ii) In {\em the two step model}, the transcription initiation consists
of two steps: first, RNAP binds to the promoter site with the on-rate
$k_{\rm on}$, and then initiates elongation with the rate $k_e$.  The
transcription initiation rate for the overall process $\Omega_0$ is
given by\cite{sneppen}
\begin{equation}
    \Omega_0 = {k_{\rm on} k_e \over k_{\rm on}+k_e}
        = {1\over \tau_{\rm on} + \tau_e}
\label{K2_0}
\end{equation}
with
\begin{equation}
  \tau_{\rm on}\equiv {1\over k_{\rm on}},
\quad
  \tau_e \equiv {1\over k_e}.
\label{tau_on-tau_e}
\end{equation}
The last expression of (\ref{K2_0}) simply shows that the average
interval of elongation initiation $1/\Omega_0$ is the sum of the two
times: $\tau_{\rm on}$, the time for RNAP to form the on-state, and
$\tau_e$, the time to start elongation in the on-state.  \vskip 1ex

iii) In {\em the three step model}, two states within the RNAP binding
state are differentiated: the one with closed DNA complex and the
other with open DNA complex.  The transition between the RNAP
unbinding state (off-state) and the RNAP binding state with closed DNA
is reversible, and characterized by the binding rate $k_b$ and
the unbinding rate $k_u$.  When RNAP is in the closed complex
state, the transition to the open state is irreversible with the rate
$k_o$.  Finally, the open complex is followed by elongation initiation
with the rate $k_e$.  This three step model of transcription
initiation was originally proposed by
Hawley-McClure\cite{hawley1982,buc}.

The three step model reduces to the two step model with the
effective on-rate $k_{\rm on}^*$ given by
\begin{equation}
 k_{\rm on}^* \equiv {k_o\over 1+k_u/k_b}
\label{k_on*}
\end{equation}
in the case where the off-state and the closed DNA binding state are
in equilibrium.
This is fulfilled when the initial reversible process of RNAP
binding/unbinding is faster than the other processes: $k_b$, $k_u \gg
k_o$, $k_e$ \cite{sneppen}.
The effective on-rate $k_{\rm on}^*$ in eq.(\ref{k_on*}) can be
understood as the open rate $k_o$ reduced by the equilibrium
expectation of being unbound.

The overall elongation rate
$\Omega_0$ for the three step model has been shown\cite{sneppen} to be
\begin{equation}
\Omega_0 = {1\over 1/k_b + 1/k_{\rm on}^* + 1/k_e}
    = {1\over \tau_b + \tau_o^* + \tau_e}
\label{K3_0}
\end{equation}
with
\begin{equation}
\tau_b\equiv {1\over k_b},
\quad
\tau_o^* \equiv {1\over k_{\rm on}^*} =
\tau_o + {k_u\over k_o}\,\tau_b ,
\quad
\tau_o \equiv {1\over k_o}
.
\label{tau_b}
\end{equation}
The time $\tau_o^*$ is the time for the system to form an open complex
after an RNAP binds to form a closed state for the first time.  It is
the sum of the two times: (i) $\tau_o$, the time to form an open complex
without unbinding, and (ii) the binding time $\tau_b$ multiplied by the
average number of times of RNAP unbindings before forming an open
complex, $k_u/k_o$ (a detailed explanation of mathematical
interpretation is given in the appendix of the supplement).  Note that
this expression holds for a general case, not limited to the case where
the two step approximation is valid.

In the above discussion, we have ignored the self-occlusion effect,
where the next RNAP cannot bind to the operator site until the previous
RNAP goes away from it.  If we include this self-occlusion effect, the
bare activity $\Omega_{\rm so}$ should be
\begin{equation}
\Omega_{\rm so} = \left[ \Omega_0^{-1}+\tau_{\rm so}\right]^{-1}
\end{equation}
with $\tau_{\rm so}$ being the time that RNAP needs to clear the promoter.

\subsection*{Transcription Factor}

For each of these models, we consider the effect of a repressive
transcription factor(TF), which we assume completely prevents RNAP from
binding while it binds to the operator site.  It is also assumed that
RNAP binding to the promoter site prevents TF from binding to the
operator site.  The binding and unbinding rates of TF are denoted by
$k_b^{\rm TF}$ and $k_u^{\rm TF}$, respectively.

\begin{figure}[tb]
\centerline{\epsfig{file=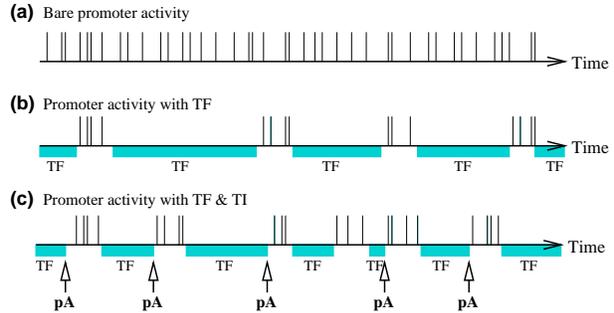,angle=0,width=8.cm,clip=}}
\caption{Schematic diagrams for the time sequence of a promoter activity
for a bare promoter (a), a promoter with TF regulation (b), and a
promoter with TF under TI (c).  The vertical lines represent the times
when transcriptions are initiated.  The shaded(cyan) intervals labeled
as TF represent the time intervals when a TF bounds to the operator
site, thus the promoter cannot initiate transcription.  Under the TF
regulation, the transcription bursts take place while a TF does not
bind.  The arrows indicate the times when interfering RNAP's from pA
arrive at pS and remove both TF and RNAP at pS; TI triggers
transcription bursts.  } \label{burst}
\end{figure}

We will study, in particular, the strong repression regime, i.e.  the
dissociation ratio $k_u^{\rm TF}/k_b^{\rm TF}$ is small.  In such a
case, TF binds for most of the time, preventing transcription
initiation, but once a TF falls off, the promoter is free to initiate a
burst of transcription elongations until another TF binds to the
operator site (Fig.\ref{burst}(b)).

\subsection*{Transcription Interference}

The effect of transcription interference(TI) on the promoter pS is
examined by exposing it to transcribing RNAP's from another promoter pA
in parallel \cite{adhya} or in convergent \cite{ward,callen}
configuration (the latter case is illustrated in Fig.\ref{fig1}(b)).
The interfering promoter pA is characterized by the transcription
initiation rate $\Omega_A$ and the initiation interval distribution
$p_A(\tau)$.  The RNAP's from pA are assumed to clear both the promoter
and the operator sites of pS (sitting duck interference) and to occlude
them while passing.  This causes bursts of transcriptions after the
interference until another TF binds(Fig.\ref{burst}(c)).

There are several additional complications related to TI.
(i) The RNAP sitting at the operator and the TF at the promoter of pS
may not simply fall off by the interfering RNAP from pA, but may block
it (roadblock effect).
(ii) Between the promoter and the operator, there should be  time
difference for the sitting duck interference and the occlusion to take
place because they extend over a certain finite size and are located at
difference places along DNA.
(iii) The interference may also take place through  collision with an
 RNAP from pA after an RNAP from pS starts elongation.
(iv) The interference between pS and pA should be  mutual, namely,
pS can also interfere in the pA activity while pA interferes with pS.

In the case where pS and pA are in a parallel configuration, the
collision effect (iii) and mutual interference (iv) do not exist.  Even
in a converging configuration, the collision effect is not significant
when the distance between pS and pA is short, i.e. the traveling time
between the two promoters is much shorter than the activity interval of
the promoters.  As for the mutual interference, the effect of pS on pA
is negligible when the activity $\Omega_A$ of pA is much larger than the
activity $\Omega$ of pS.

These effects (i)$\sim$(iv) introduce further complications in the
problem, but we are going to ignore all of them in the following.

\section*{Outline of Theory}

The quantity we are going to examine is the averaged elongation
initiation rate, or activity of pS, under the influence of TF and TI.
Under the repression by TF, a promoter initiates transcriptions in
bursts and we will see how TI can {\em activate} the promoter.  This
effect can be prominent especially when the TF repression is strong and
the time scale for TF is slow.  In this section, we outline the
theory. Detailed derivations of formulas are given in the supplement.

\subsection*{Single Promoter Property}

As tools for the analysis, we use the following two functions: (i)
$p(\tau)$, the probability distribution for time intervals between
subsequent elongation initiation events, and (ii) $C(t)$, the averaged
time-dependent rate of elongation initiation after both the promoter and
the operator sites are cleared.  We first examine $p(\tau)$ and $C(t)$
for pS without TI, but under the effect of TF.

The average elongation rate $\Omega$ without TI is the inverse of the
average elongation interval, thus it is related with $p(\tau)$ as
\begin{equation}
\Omega =  \left[ \int_0^\infty p(\tau)\tau\,  d\tau \right]^{-1}.
\end{equation}

The time dependent elongation rate $C(t)$ is actually a correlation
function of elongation initiations without TI because it can be regarded
as a probability density of initiation at the time $t$ provided that
there was an initiation at $t=0$. This can be directly calculated from
$p(\tau)$.  For large $t$, $C(t)$ approaches the promoter strength
$\Omega$,
\begin{equation}
\Omega =   \lim_{t\to\infty}C(t)
\end{equation}
because the effect of the initiation at $t=0$ lasts only a finite time.

\subsection*{Transcription Interference}

Now, we consider TI.  Under the
influence of interfering promoter pA, the promoter pS and its operator
site are assumed to be cleared every time an RNAP from pA passes, and
the activity of pS will change as $C(t)$ after that.  Thus the time
averaged activity during the interval of length $\tau$ is given by
\begin{equation}
   {1\over\tau}\int_0^{\tau-\tau_{\rm occ}}\hskip -1ex C(t)\, dt ,
\label{ave-tau}
\end{equation}
where we have included the occlusion time $\tau_{\rm occ}$.  The
occlusion time $\tau_{\rm occ}(=1 \sim 2\,{\rm sec})$ \cite{footnote}
is the time where the pS promoter cannot bind a new RNAP due to a
transcribing RNAP from pA. This effect is not included in the
correlation function $C(t)$, because the correlation function defined
here is a single promoter property.

The overall average activity of pS is the average of
eq.(\ref{ave-tau}) over the interval distribution of pA, namely,
$p_A(\tau)$. It is important to notice, however, that this average is
not with the weight $p_A(\tau)$ itself but with the weight
proportional to $p_A(\tau)\tau$ because the probability that a given
time falls in the interval of length $\tau$ is proportional to
$p_A(\tau)\tau$, not $p_A(\tau)$.  Therefore, the final expression for
the elongation rate under TI is
\begin{equation}
\Omega_{\rm TI} =
{\displaystyle
\int_{\tau_{\rm occ}}^\infty  p_A(\tau)\tau\,
  \left[ {1\over\tau}\int_0^{\tau-\tau_{\rm occ}} \hskip -1em C(t)\, dt \right]
\, d\tau
\over\displaystyle
\int_0^\infty p_A(\tau)\tau\,  d\tau }.
\label{K_TI}
\end{equation}
The occlusion effect by RNAP from pA is explicitly included as a
finite $\tau_{\rm occ}$, but the self-occlusion effect, that the RNAP
from pS blocks its own promoter site pS, should be included in the
correlation function $C(t)$ if it is considered.

%
In addition to ignoring (i) roadblock effect, (ii) time difference 
between the promoter and the operator, (iii) RNAP collision, and (iv)
mutual interference,
we will further approximate pA as Poissonian, namely,
\begin{equation}
p_A(\tau) = \Omega_A\, e^{-\Omega_A \tau},
\label{p_A}
\end{equation}
and also ignore the occlusion time by putting $\tau_{\rm occ}=0$, and the
self-occlusion effects.

\subsection*{Evaluation of $p(\tau)$ and $C(t)$}

By assuming each elementary process, such as binding, unbinding,
elongation, etc., to be a Poissonian process with a given rate, we can
obtain analytic expressions for $p(\tau)$ and $C(t)$, from which we
can calculate the overall elongation rate $\Omega$ for pS for various
situation without TI.  Using these functions, the elongation activity
under the influence of TI is estimated from eq.(\ref{K_TI}) with
$\tau_{\rm occ}=0$.

Detailed derivation of mathematical formulas is given in the
supplement.  In the following, we will describe results obtained from
those analytic expressions.

\section*{Results}

We present the numerical evaluations of our expressions for various
situations to clarify dynamical effects of TF and TI on the promoter
activity.

\begin{figure}[tb]
\centerline{\epsfig{file=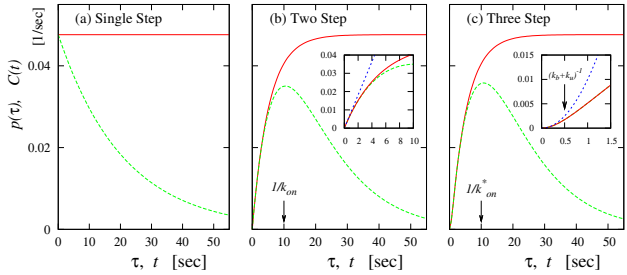,angle=0,width=8.5cm,clip=}}
\caption{\small The interval distribution $p(\tau)$(dashed green lines)
and the time dependent elongation rate $C(t)$(solid red lines) for the
single step (a), the two step (b), and the three step model (c). The
parameters for the three step model are $k_b=1 \,{\rm s}^{-1}$,
$k_u=1\,{\rm s}^{-1}$, 
$k_o=0.2\,{\rm s}^{-1}$, 
and $k_e=0.1\,{\rm
s}^{-1}$, which gives the average elongation rate
$\Omega_0=(1/k_b+1/k_{\rm on}^*+1/k_e)^{-1}=1/21\,{\rm s}^{-1}$ and the
effective on-rate $k_{\rm on}^*\equiv k_o/(1+k_u/k_b) = 0.1\,{\rm
s}^{-1}$.  The parameters for the two step model are determined so that
they behave similarly, i.e., the on-rate $k_{\rm on} = k_{\rm on}^*$ and
the elongation rate in the on-state $k_e^{(2)}$ for the two state model
given by $k_e^{(2)}=(1/\Omega_0-1/k_{\rm on}^*)^{-1}$.  The insets show
the behaviors around $t\approx 0$ with the asymptotic curves (dashed
blue lines).  } \label{fig2}
\end{figure}

\subsection*{Activity of a Bare Promoter}

Let us start by comparing the three models in a bare form, i.e.  without
TF and TI.

Fig.\ref{fig2} shows the elongation initiation interval distribution
$p(\tau)$(dashed green lines) and the time dependent activity $C(t)$
after the promoter site have been cleared by the competing
activities(solid red lines).  The parameters are chosen for the three
step model, and those for the two step and the single step models are
determined to match them with the three step model using
eqs.(\ref{k_on*}) and (\ref{K3_0}), namely, $k_{\rm on}=k_{\rm on}^*$
and $k_e$ to give the same overall activity $\Omega_0$ for all the cases.

In the single step model, the elongation initiation is Poissonian, and
the interval distribution $p(\tau)$ is a simple exponential with the
elongation rate $\Omega_0$.  As there will be no correlations between
subsequent initiations, the activity $C(t)$ is given by the constant
$\Omega_0$.

In the two step model, $p(\tau)$ and $C(t)$ rise linearly from zero as
$k_{\rm on}k_e\, t$ (the inset in Fig.\ref{fig2}(b)). This is because the
RNAP has to bind to the promoter site with the rate $k_{\rm on}$
before it initiates elongation with the rate $k_e$.  The difference
from the single step model is seen in the time scale $t \lsim
\min(1/k_{\rm on}, 1/k_e)$.  The two step model reduces to the single
step model in the case either $k_e\ll k_{\rm on}$ or $k_e\gg k_{\rm
on}$, but these two cases show quite different behaviors in reaction
to TF, as we can see in the following subsections.

In the three step model, the promoter goes through two states after RNAP
binding.  Therefore $p(\tau)$ and $C(t)$ increases initially as
$(1/2)\,k_b k_o k_e t^2$ around $t=0$ (the inset in Fig.\ref{fig2}(c)).
In the case of fast equilibration in the initial transition($k_b$, $k_u
\gg k_o$), the three step model reduces to the two step model with an
effective on-rate $k_{\rm on}^*$ given by eq.(\ref{k_on*}).

In general, the main feature of an increased number of intermediate
RNAP-promoter states causes an initial rise of $p(\tau)$
and consequently
$C(t)$ to be of increasing order in $\tau$ or $t$. Also the peak in $p(\tau)$
becomes sharper, which in principle could give a non-monotonic
behavior of $C(t)$.  For any realistic parameters, however, we find
monotonic $C(t)$ for the promoters without TF.

\subsection*{Activity of Regulated Promoter by TF}

We now consider a promoter which is regulated by a TF that acts as
repressor as illustrated in Fig.\ref{fig1}(a).  
Under strong repression by a slow binding TF, transcriptions occur in
bursts with quiescent periods of the length
\begin{equation}
\tau_{\rm TF} \equiv {1\over k_u^{\rm TF}},
\label{tau_TF}
\end{equation}
when a TF binds to the operator and suppresses the activity.  We will
see the general expressions of promoter activity $\Omega_{\rm TF}$
repressed by TF can be put in the form that allows direct interpretation
in terms of transcription burst.
We evaluate
the time-dependent activity $C(t)$ for various parameters under the
influence of TF, whose binding and unbinding rates are $k_b^{\rm
TF}=1\, {\rm s}^{-1}$ with $k_u^{\rm TF}/k_b^{\rm TF}=$ 0.1, 0.01, and
0.001.  $C(t)$'s without TF and with TF which never unbinds,
i.e. $k_u^{\rm TF}=0$, are also plotted for comparison(dashed green
lines).

\begin{figure}[tb]
\epsfig{file=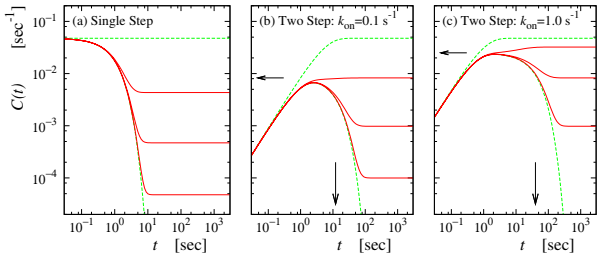,angle=0,width=8.cm,clip=}
\caption{\small The time dependent activity profile $C(t)$ with TF for
the single step (a) and the two step models with $k_{\rm on}=0.1\,{\rm
s^{-1}}$(b) and $1\,{\rm s^{-1}}$ (c).  The bare activity is one
transcription initiation per 20 sec: $\Omega_0=1/20\,{\rm s^{-1}}$. For
each case, we show five curves: the un-repressed case without TF (top
dashed green lines), the repressed cases by TF with the binding rate
$k_b^{\rm TF}=1\,{\rm s^{-1}}$ and the unbinding rate $k_u^{\rm
TF}=0.1\,{\rm s^{-1}}$ (top solid red lines), $0.01\,{\rm s^{-1}}$
(middle solid red lines), $0.001\,{\rm s^{-1}}$ (bottom solid red
lines), and with TF that never unbinds (bottom dashed green lines).
The arrows indicate $C_{\rm max}$ given by eq.(\ref{C_max}) and
$t_{\rm pl}$ given by eq.(\ref{t_pl}) for the two step models.
For the single step model(a), the RNAP activity is limited only by a
binding event once every 20 sec. For the case (b) of the two step
model, each step takes 10 sec, while in the case of (c)  where
the on-rate is fast, overall activity limited by
an elongation
initiation time of 19 sec.  }
\label{2step}
\end{figure}

\subsubsection*{Single step model}

The TF effect on the single step model is rather
straightforward. The  expression for $C(t)$ is given by
\begin{equation}
C(t) = {\Omega_0 \over k_b^{\rm TF}+ k_u^{\rm TF}}\left(
k_u^{\rm TF} + k_b^{\rm TF} e^{-(k_u^{\rm TF} + k_b^{\rm TF})t}
\right) ,
\end{equation}
which is plotted in Fig.\ref{2step}(a).  Immediately after the promoter
is cleared at $t=0$, the promoter activity recovers to the bare value
$\Omega_0$, but the initial high activity decreases as a TF binds
around $t \sim 1/k_b^{\rm TF}$. In the latter stage, the transcription
initiation is determined by the equilibrium probability of having a free
promoter, $k_u^{\rm TF}/(k_b^{\rm TF}+k_u^{\rm TF})$.  Therefore, $C(t)$
shows an exponential decrease from the initial bare activity $\Omega_0$
to the repressed level of averaged activity,
\begin{equation}
\Omega_{\rm TF} = {k_u^{\rm TF}\over k_b^{\rm TF}+k_u^{\rm TF} }\, \Omega_0 ,
\label{K1_TF-eq}
\end{equation}
for $t \gg 1/k_b^{\rm TF}$.  Note that this simple ``equilibrium
repression formula''\cite{shea} for transcription repression holds only for the
single step model.  More subtle competition comes into the problem
for the two and three step model, as we will see below.

It is interesting to see that the equilibrium formula (\ref{K1_TF-eq})
can be also put in the form
\begin{equation}
\Omega_{\rm TF} = { n_{\rm bst} \over \tau_{\rm bst} + \tau_{\rm TF} }
\label{K1_TF}
\end{equation}
with
\begin{equation}
\tau_{\rm bst}\equiv {1\over k_b^{\rm TF}},
\quad
n_{\rm bst}\equiv \Omega_0\, \tau_{\rm bst},
\label{tau1_bst}
\end{equation}
and $\tau_{\rm TF}$ defined in (\ref{tau_TF}).
This allows direct interpretation in terms of transcription burst;
$\tau_{\rm bst}$ and $n_{\rm bst}$ are the typical time scale and the
number of transcriptions, respectively, of a single transcription burst,
and $\tau_{\rm TF}$ is the typical time scale of the quiescent period
between the bursts with TF bound to the operator.  The expression
(\ref{K1_TF}) represents that the average promoter activity $\Omega_{\rm
TF}$ is given by the number of transcriptions in a burst $n_{\rm bst}$
divided by the time interval between the consecutive bursts, $\tau_{\rm
bst}+\tau_{\rm TF}$.  Note that the expression (\ref{K1_TF}) itself is
valid in general case and not limited to the case where the
transcriptions occurs in burst, namely, the promoter is strongly
repressed by a slow binding TF.

\subsubsection*{Two step model}

The situation is a little more complicated for the two step model.  In
Fig.\ref{2step}(b) and (c), two cases are shown: one with $k_{\rm
on}=0.1\, {\rm s^{-1}}$ and the other with $k_{\rm on}=1\, {\rm
s^{-1}}$; In the first case, the time scales of the two transitions, the
on-rate and the elongation rate, are same, but in the second case, the
on-rate is much faster than the elongation rate.  The elongation rate
$k_e$ are chosen to give the same bare activity $\Omega_0$ for the two
cases.

The general behavior of $C(t)$ is that (i) first it increases as $k_{\rm
on} k_e t$ until TF starts binding, (ii) then it reaches a plateau
value, and (iii) finally it goes to the  steady
activity $\Omega_{\rm TF}$ averaged over long time.

The time averaged activity with TF is given by
\begin{equation}
\Omega_{\rm TF}  =
{k_u^{\rm TF}\over [k_e/(k_{\rm on}+k_e)]\, k_b^{\rm TF} + k_u^{\rm TF} }\,\Omega_0,
\label{K2_TF-1}
\end{equation}
with $\Omega_0$ being the bare activity of the two step model
(\ref{K2_0}).  Note that the repression factor by TF, i.e. $\Omega_{\rm
TF}/\Omega_0$, is given by the ``equilibrium formula'' (\ref{K1_TF-eq})
only when $k_e\gg k_{\rm on}$.  In the other limit, TF cannot repress
the promoter as one might expect from the dissociation constant of TF,
$k_u^{\rm TF}/k_b^{\rm TF}$.

This time averaged activity (\ref{K2_TF-1}) 
can be also expressed in the same form with eq.(\ref{K1_TF}),
\begin{equation}
\Omega_{\rm TF} = { n_{\rm bst} \over \tau_{\rm bst} + \tau_{\rm TF} },
\label{K2_TF}
\end{equation}
but $n_{\rm bst}$ and $\tau_{\rm bst}$ are given by
\begin{equation}
n_{\rm bst}\equiv {k_{\rm on}\over k_b^{\rm TF}},
\quad
\tau_{\rm bst}\equiv {1\over k_b^{\rm TF}}+n_{\rm bst}{1\over k_e},
\label{n2_bst}
\end{equation}
Here, $n_{\rm bst}$ can be understood as the number of transcriptions in
a burst before a TF binds to the operator because $k_{\rm on}/k_b^{\rm
TF}$ is the ``winning ratio'' of RNAP to TF for binding.  The bursting
time $\tau_{\rm bst}$ is the sum of the binding time of TF, $1/k_b^{\rm
TF}$, and the elongation time, $1/k_e$, multiplied by the number of
transcriptions.  Again, this expressions is valid in general case
although it is interpreted best in the bursting situation.

In the strong repression limit where the bursting time is
negligible compared with the quiescent time, we have
\begin{equation}
\Omega_{\rm TF}
 \approx
{ k_{\rm on} \over k_b^{\rm TF}} \cdot
k_u^{\rm TF}
\qquad
\mbox{when }
\tau_{\rm bst} \ll \tau_{\rm TF}.
\label{K2_TF-asympt}
\end{equation}
Note that the time averaged activity in this limit does not depend on
the elongation rate $k_e$ in the on-state.  This is because the time
scale is set by the slowest rate $k_u^{\rm TF}$.  The promoter produces
a burst of $n_{\rm bst}(=k_{\rm on}/k_b^{\rm TF})$ transcriptions while
a TF is not bound, but once a TF binds, it has to wait a time $\sim
\tau_{\rm TF}(=1/k_u^{\rm TF})$ for TF to unbind.

In the case $k_b^{\rm TF}\gg k_e \gg k_u^{\rm TF}$, the plateau
becomes a maximum;  $C(t)$ can be approximated as
\begin{eqnarray}
C(t) & \approx & {k_e k_{\rm on}\over k_b^{\rm TF} + k_{\rm on}}
\Bigl(
   e^{-k_b^{\rm TF}k_e/(k_b^{\rm TF}+k_{\rm on})\cdot t}
\nonumber \\
& & \qquad  \qquad  \qquad 
-   e^{-(k_b^{\rm TF}+k_{\rm on})t}
\Bigr)
\end{eqnarray}
for $t\lsim 1/k_e\cdot \ln(k_e/k_u^{\rm TF})$ (see the supplement).
From this expression, we can estimate the maximum value $C_{\rm max}$ as
\begin{equation}
C_{\rm max}\approx {k_{\rm on}\over k_b^{\rm TF}+k_{\rm on}}\, k_e
    = {n_{\rm bst}\over t_{\rm pl}}
\label{C_max}
\end{equation}
with the plateau time
\begin{equation}
t_{\rm pl}\equiv \Bigl(  n_{\rm bst} +1 \Bigr) \,{1\over k_e}.
\label{t_pl}
\end{equation}
for the time region
\begin{equation}
{1\over k_b^{\rm TF}+k_{\rm on}}\lsim t \lsim t_{\rm pl}.
\end{equation}

The time dependent activity $C(t)$ shows maximum after the promoter site
is clarified. The maximum value (\ref{C_max}) can be understood as $k_e$
times branching probability to the on-state $k_{\rm on}/(k_b^{\rm
TF}+k_{\rm on})$.  This and eq.(\ref{K2_TF-asympt}) show that the
promoter repression by TF is determined by the competition between TF
and RNAP for binding to DNA, namely, between the binding rate $k_b^{\rm
TF}$ and $k_{\rm on}$.  Therefore, even if the bare activity is the
same, the repression by a TF can be quite different. This can be seen in
Fig.\ref{2step}: $k_{\rm on}=0.1\, {\rm s}^{-1}$ (b) and $1\, {\rm
s}^{-1}$ (c) with the same $\Omega_0=0.05\, {\rm s}^{-1}$.  The
repression in Fig.\ref{2step}(c) is about 10 times weaker than that in
(b), because $k_{\rm on}$ is 10 times faster.

After TF falls off from the operator site, the promoter produces a burst
of $n_{\rm bst}(=k_{\rm on}/k_b^{\rm TF})$ transcriptions on average
before another TF binds.  
Note that $t_{\rm pl}\approx \tau_{\rm bst}$  in the case
$k_{\rm on}\gg k_b^{\rm TF}$, namely, $n_{\rm bst}\gg 1$.

\begin{figure}[tb]
\epsfig{file=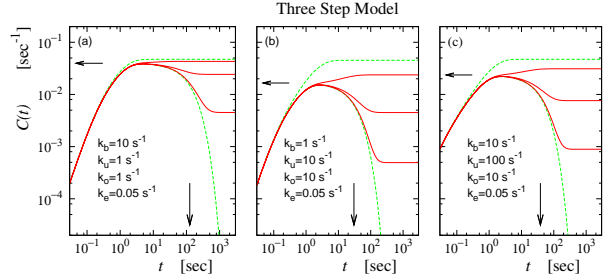,angle=0,width=8.cm,clip=}
\caption{\small
The time dependent activity profile $C(t)$ with TF for
the three step models.
For each case, we show five lines: the un-repressed case without TF (top
dashed green lines), the repressed cases by TF with the binding rate
$k_b^{\rm TF}=1\,{\rm s^{-1}}$ and the unbinding rate $k_u^{\rm
TF}=0.1\,{\rm s^{-1}}$ (top solid red lines), $0.01\,{\rm s^{-1}}$ (middle
solid red lines), $0.001\,{\rm s^{-1}}$ (bottom solid red lines), and with TF
that never unbinds (bottom dashed green lines).  The arrows indicate $C_{\rm
max}$ given by eq.(\ref{C3_max}) and $t_{\rm pl}$ given by
eq.(\ref{t3_pl}).
For all cases, the bare activity is $\Omega_0\approx 0.05\,{\rm s^{-1}}$,
but we focus on the promoters with fast open complex formation,
namely, the larger effective on-rate $k_{\rm on}^*=10/11\,{\rm
s^{-1}}$.
The case (a) corresponds with a strong closed complex binding
$k_u/k_b\ll 1$, whereas (b) and (c) deals with a weakly binding RNAP.
The difference between (b) and (c) illustrates the effect of a 10 times
faster RNAP binding rate to the promoter.
} \label{3step}
\end{figure}

\subsubsection*{Three step model}

In the full three step model, the RNAP have to pass through a closed
DNA complex state first. The transition between this closed complex
state and the off-state is reversible, but its rates
can be relatively fast compared with the transition rates of the
following steps.  The fast initial binding process tends to make
TF repression less efficient.  This has been verified by measurements on
promoters with strong RNAP binding affinity \cite{lanzer}.

The general expression for the time averaged activity with TF is again
given by
\begin{equation}
\Omega_{\rm TF} = { n_{\rm bst} \over \tau_{\rm bst} + \tau_{\rm TF} }
\label{K3_TF}
\end{equation}
with
\begin{eqnarray}
n_{\rm bst} & \equiv & {k_b\over k_b^{\rm TF}}\cdot {k_o\over k_o+k_u},
\label{n3_bst}
\\
\tau_{\rm bst} & \equiv & {1\over k_b^{\rm TF}}+n_{\rm bst}
\left({1\over k_o}+{1\over k_e}\right).
\label{tau3_bst}
\end{eqnarray}
The number of transcriptions in a burst $n_{\rm bst}$ is now given by
the winning ratio $k_b/k_b^{\rm TF}$ of RNAP to TF multiplied by the
branching ratio $k_o/(k_o+k_u)$ in the closed state to the open state.
The bursting time $\tau_{\rm bst}$ is the sum of the TF binding time
$1/k_b^{\rm TF}$ and the time needed for elongation after RNAP binding
to the promoter multiplied by the number of transcriptions.
Note that the bare activity $\Omega_0$ in eq.(\ref{K3_0}) can be
expressed as
\begin{equation}
\Omega_0 = {n_{\rm bst}\over \tau_{\rm bst}},
\end{equation}
which also holds for the other two models.

It is easy to see from eqs.(\ref{K3_0}) and (\ref{K3_TF}) that the
repression factor $\Omega_{\rm TF}/\Omega_0$ is given by the
``equilibrium formula'' (\ref{K1_TF-eq}) only when $k_e$, $k_o\gg k_b$,
$k_b^{\rm TF}$, namely, the internal time scales are negligible.
Note that the expression (\ref{K3_TF}) can be put also in Michaelis-Menten
form using (effective) dissociation constants (See Appendix).

In the strong repression limit where the bursting time is negligible,
it is easy to see that
\begin{eqnarray}
\Omega_{\rm TF} & \approx &
{n_{\rm bst}\over \tau_{\rm TF}}
\, = \,
\frac{ k_o}{k_u+k_o} \cdot {k_b\over k_b^{\rm TF}}\cdot k_u^{\rm TF}
\nonumber \\
& & \qquad  \qquad 
\mbox{ when }
\tau_{\rm bst}\ll \tau_{\rm TF}
\label{K3_TF_limit}
\end{eqnarray}
from eqs.(\ref{K3_TF}) and (\ref{n3_bst}). 
This reduces to eq.(\ref{K2_TF-asympt}) with $k_{\rm on}$ replaced by
$k_{\rm on}^*$ of eq.(\ref{k_on*}), in
the case of a weakly bound closed complex ($k_u \gg k_b,\, k_o$),
because $k_{\rm on}^* \approx k_ok_b/k_u$ in this limit.

Fig.\ref{3step} shows the time dependent activity profiles for three
promoters whose bare activities are similar but with different closed
complex formation transition rates. The first two cases, (a) and (b),
are for the same $k_{\rm on}^*=0.909\, {\rm s^{-1}}$, but for the last
case (c) $k_{\rm on}^*=0.0545\,{\rm s^{-1}}$.  One see that the
promoters respond differently to repression by a TF.  The arrows show
the maximum value $C_{\rm max}$ of eq.(\ref{C_max})  with the plateau
time 
\begin{equation}
t_{\rm pl}\equiv \Bigl(  n_{\rm bst} +1 \Bigr) \,
                 \left( {1\over k_o}+{1\over k_e} \right)
\label{t3_pl}
\end{equation}
and $n_{\rm bst}$ for the three step model.

\subsubsection*{Schematic Description for Time-Dependent Activity}

\begin{figure}[tb]
\centerline{\epsfig{file=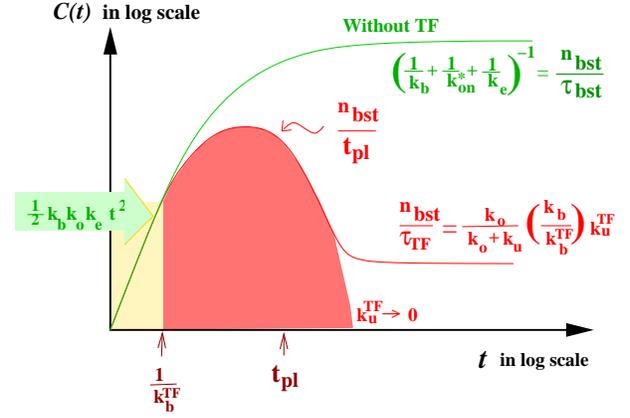,angle=0,width=8cm,clip=}}
\caption{\small Schematic activity profile of a promoter, with and
without a TF that acts as a repressor by occluding the promoter site.
Without repressor the promoter activity is set by the time that the
promoter takes to pass through the three steps to initiate elongation,
whereas a repressor reduces the final promoter activity to the extent
proportional to the dissociation ratio $k_u^{TF}/k_b^{TF}$.  Shortly
after the promoter clearance, the activity recovers to reach the maximum
value $n_{\rm bst}/t_{\rm pl}$ until $t\lsim t_{\rm pl}$.  This can be
much higher than the steady activity $n_{\rm bst}/\tau_{\rm TF}$ when
$t_{\rm tl}\ll \tau_{\rm TF}$, i.e. the promoter is strongly repressed
($\tau_{\rm TF}\gg t_{\rm pl}, \tau_{\rm bst}$) by a slow binding
TF($k_b^{\rm TF}\ll k_b$ or $n_{\rm bst}\gg 1$).  } \label{fig-explain}
\end{figure}

With all these results, Fig.\ref{fig-explain} summarizes the behavior of
time dependent activity $C(t)$ for the promoter with fast initial
binding/unbinding under the strong but slow TF repression:
\begin{equation}
(k_u,\, k_b) \, \gsim ( k_b^{\rm TF},\, k_o,\, k_e) \gg k_u^{\rm TF},
\label{app_regime}
\end{equation}
where we have a typical bursting of transcriptions with
\begin{equation}
n_{\rm bst}\gsim 1, \quad
\tau_{\rm TF}\gg t_{\rm pl} \approx \tau_{\rm bst}.
\end{equation}

After the clarification of the promoter and operator sites,
the activity increases initially as
\begin{equation}
C(t) \approx {1\over 2}k_b k_o k_e t^2,
\qquad
\mbox{for }
t\lsim {1\over k_b^{\rm TF}}
\end{equation}
until TF starts binding.

Then, it reaches the (maximum) plateau value:
\begin{equation}
C_{\rm max} \approx {n_{\rm bst}\over t_{\rm pl}}
  = {n_{\rm bst}\over n_{\rm bst}+1}\cdot {1\over 1/k_o + 1/k_e}
\quad
\mbox{for } t\lsim t_{\rm pl} .
\label{C3_max}
\end{equation}

Finally, $C(t)$ diminishes down to the long time averaged steady value
with TF,
\begin{equation}
\Omega_{\rm TF} ={n_{\rm bst}\over \tau_{\rm bst}+\tau_{\rm TF}}
\approx  {n_{\rm bst}\over \tau_{\rm TF}}
\label{K_TF}
\end{equation}

From eqs.(\ref{C3_max}) and (\ref{K_TF}),
the enhancement factor $f_{\rm enh}$ that the promoter can be
activated after the clearance of the site is given by
\begin{equation}
f_{\rm enh} \equiv {C_{\rm max}\over \Omega_{\rm TF}} 
  \approx { \tau_{\rm bst}+\tau_{\rm TF} \over t_{\rm pl}}
  \approx {\tau_{\rm TF}\over t_{\rm pl}}
\label{f_enh}
\end{equation}
This expression formalizes our original discussion that one obtain large
relative peak activity when TF repression is strong, $(\tau_{\rm bst},\,
t_{\rm pl}) \ll \tau_{\rm TF}$, but slow $k_b^{\rm TF}\ll k_b$, namely,
$n_{\rm bst}\gg 1$.  The promoters with shorter ``internal time''
$1/k_o+1/k_e$ have larger relative peak activity, and therefore they
will be more prone to de-repression by TI.

\subsection*{Interfering with Regulated Promoter Activity}

We now consider the interfering promoters pS and pA where pA is
relatively strong in comparison with pS, and pS is strongly repressed
by a slow TF.  In this case, the average activity $\Omega_{\rm TI}$ is given
by eq.(\ref{K_TI}), using the time-dependent activity $C(t)$ without
TI and the elongation interval distribution $p_A(\tau)$ of pA.

In the following, we ignore the occlusion time $\tau_{\rm occ}$ by
RNAP from pA; This should not be bad for the promoter whose
activity is of order or less than 0.1 s$^{-1}$, but may not be so
good for a more active promoter.  For $p_A(\tau)$, we will use the
exponential distribution (\ref{p_A}), which corresponds to the single
step Poissonian promoter pA.

The expression for $\Omega_{\rm TI}$ of (\ref{K_TI}) basically gives the
average of $C(t)$ over the typical time scale of pA, which is
$\Omega_A^{-1}$.  Therefore, if you look at $\Omega_{\rm TF}$ as a function of
$\Omega_A$, then $\Omega_{\rm TI}$ would show a maximum around $\Omega_A \sim
1/t_{\rm max}$ in the case that $C(t)$ has a maximum around $t\approx
t_{\rm max}$.

\paragraph{Two step model:} 
we can obtain the explicit expression
 for $\Omega_{\rm TI}$, which can be approximated as
\begin{equation}
\Omega_{\rm TI}  \approx 
{n_{\rm bst}\over t_{\rm pl}+\Omega_A^{-1}}\times
{k_b^{\rm TF}+k_{\rm on}\over k_b^{\rm TF}+k_{\rm on}+\Omega_A}
\label{K2_TI}
\end{equation}
for $\Omega_A\gg k_u^{\rm TF}$ in the regime $k_b^{\rm TF}\gg k_e \gg
k_u^{\rm TF}$.  Here, $n_{\rm bst}$ is the number of transcriptions in a
burst defined in eq.(\ref{n2_bst}), and $t_{\rm pl}$ is the plateau time
(\ref{t_pl}).
This expresses the activity of the de-repressed two-step promoter in
terms of the product of two factors; the first factor corresponds to
the averaged activity for the bursting whose interval is given by
$t_{\rm pl}+\Omega_A^{-1}$.  This factor represents the de-repression by
the interference through removal of TF by RNAP from pA before it
dissociates by itself.  The second factor represents the suppression by
removing the open complex, i.e. sitting duck interference.

The expression (\ref{K2_TI}) shows a maximum
\begin{equation}
\Omega_{\rm TI, max} \approx  {n_{\rm bst}\over t_{\rm pl}}
\quad
\mbox{at }
\Omega_A \approx \sqrt{k_b^{\rm TF}+k_{\rm on}\over t_{\rm pl}}
\label{K2_TI_max}
\end{equation}
which corresponds to eq.(\ref{C_max}).
{\em The promoter activity is actually increased by the transcription
interference.}  The enhancement factor, or the ratio $\Omega_{\rm TI,
max}/\Omega_{\rm TF}$, is the same as in eq.(\ref{f_enh}).

\begin{figure}[tb]
\epsfig{file=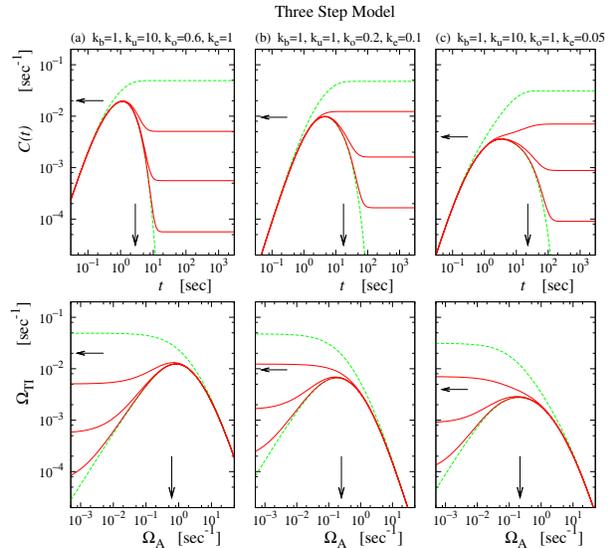,angle=0,width=8.cm,clip=}
\caption{\small
The time dependent activity profile $C(t)$ with TF
(upper plots) and the average activity $\Omega_{\rm TI}$ with TF and TI
vs. $\Omega_A$ (lower plots) for the three step models.
For each case, we show five lines: the un-repressed case without TF (top
dashed green lines), the repressed cases by TF with the binding rate
$k_b^{\rm TF}=1\,{\rm s^{-1}}$ and the unbinding rate $k_u^{\rm
TF}=0.1\,{\rm s^{-1}}$ (top solid red lines), $0.01\,{\rm s^{-1}}$
(middle solid red lines), $0.001\,{\rm s^{-1}}$ (bottom solid red
lines), and with TF that never unbinds (bottom dashed green lines).  The
arrows indicate $C_{\rm max}$ given by eq.(\ref{C3_max}) and $t_{\rm
pl}$ given by eq.(\ref{t3_pl}) for the upper plots, or $\Omega_{\rm
TI,max}$ and $\Omega_A$ by eq.(\ref{K2_TI_max}) with $k_{\rm on}$
replaced by $k_{\rm on}^*$ and $n_{\rm bst}$ and $t_{\rm pl}$ replaced
by ones of the three step model for the lower plots.
Notice that the activity
at a given level of pA activity, reflects the average activity
of the promoter up to a cut-off time of $1/\Omega_A$.
}
\label{PA-fig}
\end{figure}

\paragraph{Three step model:}
We cannot write down a compact expression, but Fig.\ref{PA-fig} shows
$\Omega_{\rm TI}$ vs. $\Omega_A$ (lower panels) along with the
corresponding $C(t)$ (upper panels) for the three step model with
different values of parameters.  One can see the correspondence between
the upper and lower panels: The activity as function of $\Omega_A$
approximately resembles the activity profiles a time $t\sim 1/\Omega_A$
plotted in corresponding upper panels.  Notice also that the potential
activation by a convergent promoter is largest for large $k_o$ and
$k_e$, as expected from eq.(\ref{f_enh}).  Finally, the relative effect
of de-repression can be very large, in the case of very slow
dissociation rate for the transcription factor.

\section*{Summary and Discussions}

We have presented a mathematical framework that expresses the dynamics
of a promoter in the Hawley-McClure model.  The formalism opens for
discussion on the promoter activity with transcription factors (TF)
and transcriptional interference (TI) by an interfering promoter, and
allows us to deal with the interplay among these elements.

In our formalism, the activity of a single promoter is characterized
by the correlation function $C(t)$, which represents the averaged time
dependent activity after the transcription initiation at $t=0$.  Any
modifications to the activity of the single promoter, such as that by
TF, are taken into account through the correlation function $C(t)$.  On
the other hand, the effects from an interfering promoter punctuates the
promoter/operator activity with the time scale of transcription
initiation from the interfering promoter.  This is represented by the
expression (\ref{K_TI}).

We have studied the effects of a repressive TF on the promoter activity.
The general expressions for the promoter activity can be put in the form
that is associated with the transcription burst, namely, a burst of
$n_{\rm bst}$ transcriptions during the bursting time $\tau_{\rm bst}$
followed by a quiescent period of the length $\tau_{\rm TF}$.  This is
actually what happens in the case of a strongly repressed promoter by a
slow binding TF.  It should be noted that the ``equilibrium formula''
(\ref{K1_TF-eq}) for the promoter activity repressed by TF is not valid
unless the time scales of internal processes are negligible compared
with binding/unbinding times of RNAP, because a TF competes with RNAP
for binding to DNA.

Under the transcription interference (TI) considered in the present
work, an interfering RNAP simply clears both the promoter and operator
sites.  If the promoter is strongly repressed by a TF, such interference
is most likely to relieve the promoter out of repression, and
interrupts the quiescent period to shorten to $1/\Omega_A$ when
$\Omega_A > k_u^{TF}$.

\subsection*{Experimental Observations}

Let us discuss experimental relevance of our theoretical results for
transcription burst and its modification by transcription interference.

\paragraph{Transcription Bursts:}

Experimentally, bunched promoter activities have been seen in several
eucaryotic systems involving TF's \cite{chubb2006,raj2006}, and they
have been interpreted to occur in response to transformation in
heterochromatin states, or as a result of a promoter approaching
transcription factories \cite{cook1999}.  For procaryotes, bunched
activities have only been observed for the promoter $P_{\rm lac/ara}$ 
under the fully induced condition\cite{golding}, and has been
interpreted without TF\cite{mitarai}.


On the other hand, transcription bursts induced by activators have been
examined in models and experiments on the yeast GAL1-promoter
\cite{blake06}, considering transcription activation by the TATA-binding
protein.  The operator position has been also shown to influence
``bunchiness'' of a promoter\cite{murphy}.

To the best of our knowledge, transcription bursts due to repressor as
are analyzed in the present paper have not been observed yet
experimentally.


\paragraph*{Transcription Interference:}

We have, at present, no direct experimental evidences for the possibility
of de-repression by transcriptional interference(TI).
Its biological relevance, however, could be widespread in
phage and {\it E. coli}; Convergent promoters is a common regulatory
motif for all temperate phages that has a CII like protein, and about 100
examples of convergent promoters have been also found in {\it E. coli}
\cite{sneppen}.

To show how TI with de-repression could help us to understand a
biological system, let us discuss the hyp-mutant of lambda; This system
is intriguing because of its high production of Cro in the lysogeny and
its enhanced immunity against infection of other
lambdas\cite{eisen,georgopoulos}.
Its DNA configuration resembles that in Fig.\ref{fig1}(b) and the
parameters in Fig.\ref{PA-fig} are matched to this system.  Therefore,
the maximal repression case there corresponds to the case of the
promoter PR in lambda repressed by the factor of about 500 due to CI
\cite{revet,dodd2004,dodd-pc}.  The strength of hyp-PRE is not known,
but Fig.\ref{PA-fig} suggests that PR in lysogen could be de-repressed
by the factor 10$\sim$30 due to TI from hyp-PRE, provided that open
complex formation is fast and that CI binds to OR relatively slowly.
This could explain, at least, a part of the large amount of Cro found in
the lysogeny of the hyp-mutant.

Another example is the $O_{R3}O_{R2}O_{R3}$ mutant, which has been also
found to show stable lysogens\cite{little}, even though it is expected
to be producing Cro 10$\sim$30 times more than a normal
lambda\cite{sneppenbook}, as in the case of the hyp-mutant.  Such
similarity, i.e. the stable lysogens under the high production of Cro,
between the $O_{R3}O_{R2}O_{R3}$ and the hyp-mutant leads us to
speculate that the remarkable robustness of the lysogens\cite{aurell} of
these phages should be rooted in the same unknown mechanism.

\subsection*{Experimental Proposal}

Burst activity should be most directly monitored by a real time
observation, but also can be examined quantitatively from the number
distribution of mRNA in a cell.  This may be obtained if one can take
snapshots of an assembly of cells from which the number of mRNA
contained in each cell can be counted.  The reaction rate constants for
RNAP and TF should be able to be estimated from the mRNA distribution.

For example, from the distribution one can calculate the Fano factor
$\nu$, which is the ratio of the variance to the average
\begin{equation}
\nu\equiv {\left<(n-\left<n\right>)^2\right>\over \left<n\right>}
\end{equation}
with $n$ being the number of mRNA in a cell.  This Fano factor can be
directly compared with our estimate of the number of transcriptions in a
burst $n_{\rm bst}$; In the bursting situation with $n_{\rm bst}\gg 1$, the
Fano factor should be given by $n_{\rm bst}$ if the quiescent periods
follow Poissonian process and are much longer than the bursting periods:
\begin{equation}
\nu \sim n_{\rm bst}={k_b\over k_b^{\rm TF}}\cdot{k_o\over k_o+k_u}
\qquad
\mbox{for }
\tau_{\rm bst}\ll  \tau_{\rm TF},
\end{equation}
but is smaller than that if the burstings are not separated well enough:
\begin{equation}
\nu < n_{\rm bst}
\qquad
\mbox{for }
\tau_{\rm bst}\lsim  \tau_{\rm TF}.
\end{equation}
In the case $n_{\rm bst}\ll 1$, we would have $\nu \sim 1$ because each
elongation initiation follows the Poissonian process.  The full
information of the distribution allows us more detailed comparison with
our analysis.

Another experiment we can propose is to construct DNA with a promoter
exposed to a library of interfering promoters with varying strength
$\Omega_A$, preferably in a parallel configuration to avoid RNAP collisions.
Suppose the promoter is highly repressed by TF with unknown parameters.
By examining how the promoter is de-repressed by the interfering
promoters, the off-rate $k_u^{TF}$ of TF can be estimated as the lower
limit of $\Omega_A$ that de-represses the promoter.

\subsection*{Simplifications in the Present Treatment}

Before concluding, let us discuss some of the effects we have
ignored in the present treatment.

\paragraph{Roadblock:}
In our analysis of TI, we have assumed that RNAP always displaces TF
without roadblock effect, but it is known that some TF's are roadblocks
to RNAP.  Roadblocks are most commonly reported in in-vitro experiments
\cite{pavco,izban,reines},
%
%
whereas presence of elongation factors often allow RNAP to pass the
roadblock in in-vivo situations \cite{reines, toulme}.
%
Reports on in-vivo roadblocks is at present limited to the transcription
factor {\it LacI} and the restriction enzyme EcoRI \cite{epshtein}.
It has been reported that some roadblocks may be translocated, being
pushed by two or more RNAP's\cite{epshtein}.  If two consecutive RNAP's
are required to dislocate a TF, the activity of interfering promoter
$\Omega_A$ in eq.(\ref{K2_TI}) should be replaced by the effective
activity, which is half of the original activity, $\Omega_A/2$.  This
reduction factor 1/2 should be further reduced in the case 
where a blocked RNAP may fall off before the second one arrives to give
a push.

Another possibility that RNAP does not remove TF is that RNAP just
passes TF without displacing it; The repressor would simply not leave
the vicinity of the operator, and thus maintains its function until it
falls off by itself.  This kind of situation has been actually observed
when an RNAP reads through a nucleosome, displacing only parts of the
histone complex\cite{kireeva,walter}.
For some TF, one could imagine mixed situations, where TF is displaced
but remains in physical proximity during the RNAP passage.

TF's such as CI in phage 186 \cite{callen} and CI in $\lambda$ on OR
\cite{dodd-pc} do not act as roadblocks but are removed, which, we
presume, are more common situations.

\paragraph{Time difference for the promoter and the operator:}
The interfering RNAP's clear/occlude the promoter pS first and then the
operator in the convergent configuration, and the other way around in
the parallel configuration.  The time difference of the effects for the
two sites depends on the distance between the two sites.  If the binding
times of TF or RNAP are comparable or shorter than this time difference,
we have to take this into account, which makes the situation favorable
to the promoter(operator) in the convergent(parallel) configuration.

\paragraph{RNAP collision:}
The RNAP from pS may be removed, even after it starts elongating, by
colliding with the RNAP from pA.  This effect is profound particularly
when the distance between pS and pA is large.  It has been found that
the collision effect becomes substantial for convergent promoters with
the pS-pA distance being of the order of $v/(2 \Omega_A)$, where
$v$($\sim 50$ bp/sec) is the transcription elongation
speed\cite{sneppen}.  For the parallel configuration of promoters, the
collision effect does not exist.

\paragraph{Occlusion time:}
The occlusion time $\tau_{\rm occ}$, the time that the promoter pS is
occluded by passing the RNAP from pA, was neglected.  This has been also
considered in \cite{sneppen} and found that pS is influenced
substantially by occlusion only when the activity of pA is stronger than
0.1 s$^{-1}$.

\paragraph{Mutual interference:}
In the convergent configuration of promoters, not only pA interferes
with pS, but also pS interferes with pA.  Such mutual interference effects
are likely to be important in switching mechanisms between equally
strong convergent promoters, such as the convergent promoters PR and PRE
of lambda phage in the early stages of infection.  Full analytical
treatment on the mutual interference is not easy in general case, but
stochastic simulations \cite{sneppen} and the four-world approximation
analysis \cite{dodd2007} have been performed.  In the present analysis,
we consider the highly repressed promoter pS, thus the interference of
pS on pA should be negligible.  In the case of the parallel
configuration, this effect does not exist.

\vskip 2ex

\noindent {\bf Acknowledgment:} We very much like to thank Ian Dodd,
Alexandra Ahlgren-Berg, and Adam Palmer for discussions on transcription
interference, and Harvey Eisen for suggesting that transcribing RNAP
from the hyp-PRE promoter may reduce CI repression of PR in the
hyp-mutant of phage lambda.  We thank for financial support from the
Danish National Research Foundation through the Center for Models of
Life.

\appendix
\section*{Appendix: Promoter Activity in Michaelis-Menten Form}

Since the process of transcription initiation can be regarded as an
enzyme reaction, our results for the averaged promoter activity
in the three step model can be put in the form of Michaelis-Menten
kinetics.

Let us start by the bare activity without TF.  The binding rate $k_b$ of
RNAP should be proportional to the density of RNAP,
\begin{equation}
k_b \equiv [{\rm RNAP}] \kappa_b
\end{equation}
with a reaction constant $\kappa_b$.  Then, the bare activity
(\ref{K3_0}) can be written as
\begin{equation}
\Omega_0 = {[{\rm RNAP}]/K^*_{\rm RNAP} \over 1+ [{\rm RNAP}]/K^*_{\rm RNAP}}
       \, \Omega_0^{\rm max}
\label{K3_0-MM}
\end{equation}
with the maximum activity
\begin{equation}
\Omega_0^{\rm max} \equiv {k_o k_e \over k_o + k_e},
\end{equation}
and the effective dissociation constant for RNAP
\begin{equation}
K^*_{\rm RNAP}\equiv
    \left( 1+{k_u\over k_o}\right) {\Omega_0^{\rm max}\over \kappa_b} .
\end{equation}


TF has been introduced as a competitive inhibitor
in our model.  Its binding rate can be expressed as
\begin{equation}
k_b^{\rm TF} \equiv [{\rm TF}] \kappa_b^{\rm TF},
\end{equation}
with the TF density [{\rm TF}] and the reaction constant
$\kappa_b^{\rm TF}$, then the dissociation constant for TF is given by
\begin{equation}
 K_{\rm TF} \equiv {k_u^{\rm TF}\over  \kappa_b^{\rm TF} }.
\end{equation}
With these parameters, the expression (\ref{K3_TF}) for the averaged activity
with TF is written as
\begin{equation}
\Omega_{\rm TF} =
{\displaystyle
{[{\rm RNAP}]/ K^*_{\rm RNAP}} \over \displaystyle
1+ {[{\rm TF}]/  K_{\rm TF}} + {[{\rm RNAP}]/ K^*_{\rm RNAP}}
}\, \Omega_0^{\rm max},
\end{equation}
which is in the standard form of
Michaelis-Menten kinetics with a competitive inhibitor.

\bibliography{references}

\begin{thebibliography}{46}
\bibitem[foo()]{footnote}
$\tau_{occ}=(r+\ell)/v$ is the time needed to for a RNAP from pA totranscribe
  acros the pS promoter (length $r+\ell\sim 105bp$ and speed $v$ estimated to
  be 50bp/sec \cite{sneppen}).
\end{thebibliography}

\onecolumn

\def\lsim{\mathrel{\rlap{\lower4pt\hbox{\hskip1pt$\sim$}}
    \raise1pt\hbox{$<$}}}                
\def\gsim{\mathrel{\rlap{\lower4pt\hbox{\hskip1pt$\sim$}}
    \raise1pt\hbox{$>$}}}                


\setcounter{page}{1}
\setcounter{equation}{0}
\renewcommand{\thesection}{\Roman{section}.}
\renewcommand{\thesubsection}{\Alph{subsection}.}
\setcounter{section}{0}

\begin{center}
\large
\title{\large \bf Supplement for \\
``Dynamical Analysis on Gene Activity \\
in the Presence of Repressors and an Interfering Promoter''}
\vskip 4ex

\author{\bf
{\bf Hiizu Nakanishi $^{1,2}$}, 
{\bf Namiko Mitarai $^2$}, and {\bf Kim Sneppen $^1$}
}
\vskip 1ex

{\it
{\rm $^1$} Niels Bohr Institute , Blegdamsvej 17, Dk 2100, Copenhagen, Denmark
 \\
{\rm $^2$} Department of Physics, Kyushu University 33, Fukuoka 812-8582, Japan
} 

\vskip 1ex

(\date{\today})
\end{center}
\vskip 3ex

\begin{abstract}
Detailed derivations for the mathematical expressions in the text are
given.
\end{abstract}
\vskip 3ex

\section{Elongation Initiation Interval and Correlation}

Let $C(t)$ be the time-dependent activity after the clearance of both
the promoter and the operator.  Then it can also be regarded as a
correlation function of the transcription initiation, and it is related to
the initiation interval distribution $p(\tau)$ as
\begin{eqnarray}
\lefteqn{
C(t)  =  p(t) + 
\int_0^\infty  d\tau_1 \int_0^\infty  d\tau_2 \,
        \delta(t-\tau_1-\tau_2)\,  p(\tau_1)p(\tau_2)}
\nonumber \\
& & \quad
+\int_0^\infty  d\tau_1 \int_0^\infty  d\tau_2 \int_0^\infty d\tau_3 \,
        \delta(t-\tau_1-\tau_2-\tau_3)\, p(\tau_1)p(\tau_2) p(\tau_3)
\nonumber \\
& & \qquad +
 \cdots .
\end{eqnarray}
Since each term in the right hand side is a convolution of $p(\tau)$,
the Laplace transformation
\begin{equation}
\tilde C(s)\equiv \int_0^\infty C(t) e^{-st}dt
\end{equation}
can be obtained easily as a sum of geometrical series;
\begin{equation}
\tilde C(s) = \sum_{n=1}^{\infty} \tilde p(s)^n
= {\tilde p(s)\over 1-\tilde p(s)}
\label{S-C-s}
\end{equation}
with $\tilde p(s)$ being the Laplace transform of $p(\tau)$.

\section{Bare Promoters}

In this section, the explicit expressions for $p(\tau)$ and $C(t)$ for
a bare promoter of each model are derived within the approximation
that the self-occlusion effect is ignored.

\subsection{Single step model}
For the single step model, the elongation initiation is a simple
Poissonian process with the rate $\Omega_0$, thus we have
\begin{equation}
p(\tau) = \Omega_0 e^{-\Omega_0\tau}, \qquad
\tilde p(s) = {\Omega_0\over s+\Omega_0},
\end{equation}
and
\begin{equation}
\tilde C(s)= {\Omega_0\over s},\qquad
C(t) = \Omega_0 .
\label{S-1step-C}
\end{equation}

\subsection{Two step model}
In the two step model, each elongation interval consists of an
off-state period and an on-state period, whose length distributions,
$p_{\rm off}(\tau_{\rm off})$ and $p_{\rm on}(\tau_{\rm on})$, are
Poissonian given by
\begin{equation}
p_{\rm off}(\tau_{\rm off})=k_{\rm on}e^{-k_{\rm on}\tau_{\rm off}}, 
\quad \mbox{and}\quad
p_{\rm on}(\tau_{\rm on})=k_e e^{-k_e \tau_{\rm on}}, 
\end{equation}
respectively.  Since the elongation interval is the sum of the
off-period length and the on-period length, the elongation interval
distribution $p(\tau)$ for the two step model is given by
\begin{equation}
p(\tau) = \int_0^\infty d\tau_{\rm off} \int_0^\infty d\tau_{\rm on}\,
\delta(\tau-\tau_{\rm off}-\tau_{\rm on}) 
p_{\rm off}(\tau_{\rm off})p_{\rm on}(\tau_{\rm on}).
\end{equation}
Again, the right hand side is a convolution of  $p_{\rm
off}(\tau_{\rm off})$ and $p_{\rm on}(\tau_{\rm on})$, thus
in the Laplace transform, we have
\begin{equation}
\tilde p(s) = \tilde p_{\rm off}(s)\, \tilde p_{\rm on}(s)
= {k_{\rm on} k_e \over (s+k_{\rm on})(s+k_e)},
\end{equation}
which gives
\begin{equation}
p(\tau) = \left\{\begin{array}{ll}\displaystyle
k_{\rm on}k_e \, {e^{-k_e\tau}-e^{-k_{\rm on}\tau}\over k_{\rm on}-k_e}
& \mbox{for } k_{\rm on}\ne k_e
\\ \displaystyle
k_e^2 \,\tau \, e^{-k_e\tau}
& \mbox{for } k_{\rm on}= k_e
\end{array}\right. .
\end{equation}
Then, the correlation function is given by 
\begin{equation}
\tilde C(s) = {k_{\rm on}k_e\over s (s+k_{\rm on}+k_e)},
\quad
C(t) = \Omega_0  \left(  1-e^{-(k_{\rm on}+k_e)t} \right)
\label{S-2step-C}
\end{equation}
with $\Omega_0$ being the bare activity for the two step model:
\begin{equation}
\Omega_0= {k_{\rm on}k_e \over k_{\rm on}+k_e}
= {1\over \tau_{\rm on} + \tau_e};
\qquad
\tau_{\rm on}\equiv {1\over k_{\rm on}},
\quad
\tau_e \equiv {1\over k_e }.
\label{S-Omega_0-two-step}
\end{equation}
Note that the last expression simply represents that the average interval
between elongation $1/\Omega_0$ is the sum of the average waiting time to become
the on-state $\tau_{\rm on}$ and the time for the elongation $\tau_e$.

\subsection{Three step model}

In the three step model, the initial transition between the off-state
and the closed complex state is reversible, which means that the closed state
goes either back to the off-state or forward to the open complex state with
the branching ratios $k_u$ and $k_o$, or with the probabilities
\begin{equation}
p\equiv {k_u\over k_u+k_o}, \quad\mbox{and}\quad
q\equiv {k_o\over k_u+k_o}=1-p,
\end{equation}
respectively.  Therefore, the promoter may get into the closed state
many times before an RNAP starts elongation.  Let $n$ be the number of
times that the promoter gets in the closed state before elongation, then
the sequence of states and their probabilities are
\begin{equation}\begin{array}{ccc}
n & \mbox{state sequence} & \mbox{probability} \\
\hline
1 & \mbox{(off - closed) $\bullet$ open - elong} & q \\
2 & \quad\mbox{(off - closed) $\circ$ (off - closed) $\bullet$ open - elong}
  \quad &     p\,q \\
  & \cdot\cdot\cdot & \\
n & \mbox{(off - closed $\circ$)$^{n-1}$(off - closed) $\bullet$ open - elong}
  &  p^{n-1}q \\
  & \cdot\cdot\cdot & \\
\end{array},
\label{S-2step_seq}
\end{equation}
where $\circ$ and $\bullet$ represent the branching probabilities $p$ and
$q$, respectively.

Let $p_n(\tau)$ be the elongation interval distribution for the
interval during which the promoter goes through the closed state $n$
times, then it is given by a convolution of the life time distribution
of the off-state $p_{\rm off}(\tau)$, the closed state $p_{\rm
cl}(\tau)$, and the open state $p_{\rm op}(\tau)$; For example,
$p_1(\tau)$ is given by
\begin{equation}
p_1(\tau)=
\int_0^\infty d\tau_{\rm off}
\int_0^\infty d\tau_{\rm cl}
\int_0^\infty d\tau_{\rm op}\,
\delta(\tau-\tau_{\rm off}-\tau_{\rm cl}-\tau_{\rm op})
p_{\rm off}(\tau_{\rm off})p_{\rm cl}(\tau_{\rm cl})p_{\rm op}(\tau_{\rm op}).
\end{equation}
In the same way, the Laplace transform of $p_n(\tau)$ for general $n$
is given by
\begin{equation}
\tilde p_n(s) =
\Bigl( \tilde p_{\rm off}(s)\tilde p_{\rm cl}(s) \Bigr)^n \tilde p_{\rm op}(s)
\end{equation}
with
\begin{equation}
\tilde p_{\rm off}(s)={k_b\over s+k_b},\quad
\tilde p_{\rm op}(s)={k_u+k_o \over s+k_u+k_o},\quad
\tilde p_{\rm cl}(s)={k_e\over s+k_e} .
\end{equation}

The elongation interval distribution $p(\tau)$ is the average over
$p_n(\tau)$ with the probability given by (\ref{S-2step_seq}), and can
be calculated as follows;
\begin{equation}
\tilde p(s)  
= 
\sum_{n=1}^\infty \tilde p_n(s)\, p^{n-1}q
\quad =
{\tilde p_{\rm off}(s) \tilde p_{\rm cl}(s) \over 
1- \tilde p_{\rm off}(s) \tilde p_{\rm cl}(s) p} \,\,\tilde p_{\rm op}(s) \, q
\quad = 
{k_b k_o k_e \over (s+k_+)(s+k_-)(s+k_e)}
\label{S-p3-s}
\end{equation}
with
\begin{equation}
k_\pm \equiv 
   {1\over 2}\Bigl[ (k_b+k_u+k_o)\pm\sqrt{(k_b+k_u+k_o)^2-4k_b k_o} \Bigr] ,
\end{equation}
from which we obtain
\begin{equation}
p(\tau) = 
{k_bk_ok_e\over (k_e-k_+)(k_e-k_-)}e^{-k_e\tau} +
{k_bk_ok_e\over \sqrt{(k_b+k_o+k_e)^2-4k_bk_o}}
\left[ {e^{-k_-\tau}\over k_e-k_-}-{e^{-k_+\tau}\over k_e-k_+} \right].
\end{equation}

From eqs.(\ref{S-C-s}) and (\ref{S-p3-s}), we have
\begin{equation}
\tilde C(s)={k_b k_o k_e\over s(s+k_+^C)(s+k_-^C)};
\quad
k_\pm^C \equiv
{1\over 2}\left[ (k_b+k_u+k_o+k_e)\pm\sqrt{(k_b+k_u+k_o-k_e)^2-4k_bk_o} \right]
,
\end{equation}
which leads to
\begin{equation}
C(t)= \Omega_0  \left[  1-
{k_+^C e^{-k_-^Ct} - k_-^C e^{-k_+^Ct}\over k_+^C - k_-^C} 
\right]
\end{equation}
where  $\Omega_0$ is  the bare activity for the three step model:
\begin{equation}
 \Omega_0 = { k_b k_o k_e \over  k_+^C k_-^C }
\quad
= {1\over \tau_b + \tau_o^* + \tau_e};
\qquad
\tau_b \equiv {1\over k_b},
\quad
\tau_o^* \equiv {1\over k_{\rm on}^*} =
             {1\over k_o} + {k_u\over k_o}\cdot {1\over k_b},
\quad
\tau_e \equiv {1\over k_e}.
\label{S-K3_0}
\end{equation}
The average interval between elongations, $1/\Omega_0$, are the sum of the
three times: (1) $\tau_b$, the time for RNAP to bind and form the
closed complex for the first time, (2) $\tau_o^*$, the time for RNAP
to form the open complex after the first binding, and (3) $\tau_e$,
the time to start elongation.  Note that the validity of this
expression is {\em not} limited to the case within the two step
approximation, where $k_{\rm on}^*$ can be interpreted as the
effective on-rate.

The time $\tau_o^*$ consists of two parts: (i) $1/k_o$, the time to go
forward to the open state, and (ii) the re-binding time $1/k_b$ after
unbinding multiplied by the average number of unbindings
$k_u/k_o$. Mathematical derivation of this expression is given in the
appendix.  We will encounter similar expressions in the following.

\section{Regulated Promoters}

Now, we derive the expressions for $p(\tau)$ and $C(t)$ for each model
of a bare promoter in the case where the promoter is repressed by a
transcription factor (TF). 
In the case where the suppression is strong by a slow binding TF, the
transcription activity occurs in bursts.
The averaged activity is given by the long time limit $t\to\infty$ of
$C(t)$.

\subsection{Single step model}

The state sequence between elongations can be classified 
according to the number of TF bindings, and the probability and
the interval distribution for each case are obtained as
\begin{equation}\begin{array}{cccc}
n & \mbox{state sequence} & \mbox{probability} & \mbox{interval distribution} \\
\hline
0 & \mbox{off $\bullet$ elong} & q & \tilde p_{\rm off}(s) \\
1 & \mbox{(off $\circ$ TF) - off $\bullet$ elong} & p\, q &
 \tilde p_{\rm off}(s) \tilde p_{\rm TF}(s) \tilde p_{\rm off}(s) \\
2 & \mbox{(off $\circ$ TF)$^2$ - off $\bullet$ elong} & p^2 q &
 \bigl(\tilde p_{\rm off}(s) \tilde p_{\rm TF}(s)\bigr)^2
 \tilde p_{\rm off}(s) \\
& \cdot\cdot\cdot \\
n & \quad\mbox{(off $\circ$ TF)$^n$ - off $\bullet$ elong}\quad & p^n q &
 \bigl(\tilde p_{\rm off}(s) \tilde p_{\rm TF}(s)\bigr)^n
 \tilde p_{\rm off}(s) \\
& \cdot\cdot\cdot 
\end{array},
\end{equation}
where TF represents the state with TF at the operator site and
\begin{equation}
\tilde p_{\rm off}(s)\equiv {k_b^{\rm TF}+\Omega_0\over s+k_b^{\rm TF}+\Omega_0},\quad
\tilde p_{\rm TF}(s)\equiv {k_u^{\rm TF} \over s+k_u^{\rm TF}},\qquad
p\equiv {k_b^{\rm TF}\over k_b^{\rm TF}+\Omega_0}, \quad
q\equiv {\Omega_0 \over k_b^{\rm TF}+\Omega_0}.
\end{equation}
Thus, we have
\begin{eqnarray}
\tilde p(s) & = & \sum_{n=0}^\infty p^nq 
 \bigl(\tilde p_{\rm off}(s) \tilde p_{\rm TF}(s)\bigr)^n 
    \tilde p_{\rm off}(s) 
\quad
=
{\tilde p_{\rm off}(s)q \over 1-\tilde p_{\rm off}(s) \tilde p_{\rm TF}(s)p }
\quad
=
{(s+k_u^{\rm TF})\Omega_0\over (s+k_+)(s+k_-)};
\\
& &
k_\pm\equiv
{1\over 2}\Bigl[
k_b^{\rm TF}+k_u^{\rm TF}+\Omega_0 \pm
\sqrt{ (k_b^{\rm TF}+k_u^{\rm TF}+\Omega_0 )^2-4k_u^{\rm TF}\Omega_0}
\Bigr]
\end{eqnarray}
and
\begin{equation}
C(t) = {\Omega_0\over k_b^{\rm TF}+k_u^{\rm TF}}
   \Bigl( k_u^{\rm TF} + k_b^{\rm TF} e^{-(k_b^{\rm TF}+k_u^{\rm TF})t} \Bigr).
\label{S-1step-C-TF}
\end{equation}

Thus, the steady state activity $\Omega_{\rm TF}$ for the single step model
is given by
\begin{equation}
\Omega_{\rm TF} = \lim_{t\to\infty} C(t) = 
 {k_u^{\rm TF}\over k_b^{\rm TF}+k_u^{\rm TF}}\, \Omega_0 
\quad
=
{n_{\rm bst}\over \tau_{\rm bst}+\tau_{\rm TF}}
\label{S-1step-K_TF}
\end{equation}
with
\begin{equation}
\tau_{\rm bst}\equiv {1\over k_b^{\rm TF}},
\qquad
n_{\rm bst}\equiv \Omega_0 \tau_{\rm bst},
\qquad
\tau_{\rm TF}\equiv  {1\over k_u^{\rm TF}}
\, .
\label{S-n_bst-1}
\end{equation}
The last expression for $\Omega_{\rm TF}$ allows a
simple interpretation in terms of bursting activity; $\tau_{\rm bst}$
and $\tau_{\rm TF}$ are the bursting time and the quiescent time,
respectively, and $n_{\rm bst}$ is the number of transcriptions during
the bursting time.

\subsection{Two step model}

For the two step model, the state sequences, their probabilities, and
the interval distributions are 
\begin{equation}\begin{array}{cccc}
n & \mbox{state sequence} & \mbox{probability} & \mbox{interval distribution}
\\ \hline
0 & \mbox{off $\bullet$ on - elong} & q & \tilde p_{\rm off}(s) \tilde p_{\rm on}(s) \\
1 & \mbox{(off $\circ$ TF) - off $\bullet$ on - elong} & p\, q &
\tilde p_{\rm off}(s) \tilde p_{\rm TF}(s) 
  \tilde p_{\rm off}(s) \tilde p_{\rm on}(s) \\
2 & \mbox{(off $\circ$ TF)$^2$ - off $\bullet$ on - elong} & p^2 q &
 \bigl(\tilde p_{\rm off}(s) \tilde p_{\rm TF}(s)\bigr)^2
 \tilde p_{\rm off}(s)  \tilde p_{\rm on}(s) \\
& \cdot\cdot\cdot \\
n & \quad\mbox{(off $\circ$ TF)$^n$ - off $\bullet$ on - elong}\quad & p^n q &
 \bigl(\tilde p_{\rm off}(s) \tilde p_{\rm TF}(s)\bigr)^n
 \tilde p_{\rm off}(s)  \tilde p_{\rm on}(s) \\
& \cdot\cdot\cdot 
\end{array},
\end{equation}
with
\begin{equation}
\tilde p_{\rm off}(s)={k_b^{\rm TF}+k_{\rm on}\over s+k_b^{\rm TF}+k_{\rm on}},
\quad
\tilde p_{\rm TF}(s)={k_u^{\rm TF} \over s+k_u^{\rm TF}},\quad
\tilde p_{\rm on}(s)={k_e \over s+k_e},
\qquad
p\equiv {k_b^{\rm TF}\over k_b^{\rm TF}+k_{\rm on}},\quad
q\equiv {k_{\rm on}\over k_b^{\rm TF}+k_{\rm on}}.
\end{equation}

From these, we obtain
\begin{eqnarray}
p(\tau)
& = &
k_e k_{\rm on} \left[
{k_u^{\rm TF} -k_e \over (k_+ - k_e) (k_- - k_e)} e^{-k_e \tau}
+
{1\over k_+-k_-}\left(
{k_u^{\rm TF}-k_-\over k_e - k_-} e^{-k_-\tau} -
{k_u^{\rm TF}-k_+\over k_e - k_+} e^{-k_+\tau} 
\right)
\right]
\\
& &
k_\pm \equiv  {1\over 2}\left[
k_b^{\rm TF}+k_u^{\rm TF}+k_{\rm on} \pm \sqrt{
(k_b^{\rm TF}+k_u^{\rm TF}+k_{\rm on})^2 - 4k_u^{\rm TF}k_{\rm on} }
\right]
\\
C(t) & = &
 {k_e k_{\rm on} k_u^{\rm TF} \over k_+^Ck_-^C} + 
      {k_e k_{\rm on} \over k_+^C - k_-^C}
\left[
    {k_-^C - k_u^{\rm TF}\over k_-^C} e^{-k_-^Ct}-
         {k_+^C - k_u^{\rm TF}\over k_+^C} e^{-k_+^C t}
\right]
\\
& &
k_\pm^C \equiv 
{1\over 2}\left[ (k_b^{\rm TF} + k_{\rm on} + k_u^{\rm TF} + k_e) \pm
\sqrt{ (k_b^{\rm TF} + k_{\rm on} + k_u^{\rm TF} + k_e)^2 - 
           4(k_{\rm on}k_u^{\rm TF} + k_u^{\rm TF} k_e+k_b^{\rm TF} k_e)
}\right]
\label{S-2step-k_pmC}
\end{eqnarray}

The steady state activity $\Omega_{\rm TF}$ is now
\begin{equation}
\Omega_{\rm TF} 
=
\lim_{t\to\infty} C(t) 
\quad =
 {k_e k_{\rm on} k_u^{\rm TF} \over k_+^Ck_-^C}
\quad =
 {k_e k_{\rm on} \over k_{\rm on} +  k_e + (k_b^{\rm TF}/k_u^{\rm TF}) k_e } 
\quad =
{n_{\rm bst}\over \tau_{\rm bst}+\tau_{\rm TF}}
,
\label{S-2step-K_TF}
\end{equation}
with
\begin{equation}
n_{\rm bst} \equiv {k_{\rm on}\over k_b^{\rm TF}},
\qquad
\tau_{\rm bst}\equiv {1\over k_b^{\rm TF}}+n_{\rm bst}{1\over k_e} .
\label{S-n_bst-2}
\end{equation}
The number of transcriptions during a bursting period is given by the
winning ratio of RNAP to TF.

It is interesting to see that eq.(\ref{S-2step-K_TF}) can be also put in
the form,
\begin{equation}
\Omega_{\rm TF} = {1\over \tau_{\rm on}+\tau_e};
\qquad
\tau_{\rm on}({\rm TF})\equiv
    \tau_{\rm on}+{k_b^{\rm TF}\over k_{\rm on}}\tau_{\rm TF},
\quad
\tau_{\rm on} = {1\over k_{\rm on}},
\quad
\tau_e = {1\over k_e}
\, ,
\label{S-2step-K_TF-2}
\end{equation}
which is similar to the expression for the bare promoter
(\ref{S-Omega_0-two-step}).
The expression for $\tau_{\rm on}({\rm TF})$ allows a similar
interpretation with that for eqs.(\ref{S-K3_0}); 
the time to reach the on-state from the off-state,
$\tau_{\rm on}({\rm TF})$ is the sum of the two times: (1) $\tau_{\rm on}$,
the time to reach the on-state without TF binding, and (2) the TF
unbinding time, $\tau_{\rm TF}$, multiplied by the number of TF
bindings, $k_b^{\rm TF}/k_{\rm on}$, before an RNAP binds.

In the case $k_b^{\rm TF}\gg k_e \gg k_u^{\rm TF}\approx 0$, the
correlation function $C(t)$ shows a plateau.  As the lowest order
estimate, we put simply $k_u^{\rm TF}=0$, then we obtain
\begin{equation}
C(t) \approx {k_e k_{\rm on}\over k_b^{\rm TF}+k_{\rm on}}
\left( e^{-k_-^Ct}-e^{-k_+^Ct} \right) ; 
\qquad
k_+^C \approx k_b^{\rm TF}+k_{\rm on},\quad
k_-^C \approx {k_b^{\rm TF}k_e \over k_b^{\rm TF}+k_{\rm on}}.
\label{S-2step-C-TF-app}
\end{equation}
therefore, $C(t)$ behaves as
\begin{eqnarray*}\displaystyle
 C(t) &  \approx & \left\{\begin{array}{ll}
k_{\rm on}k_e t &
\mbox{for }\quad t\lsim (k_b^{\rm TF}+k_{\rm on})^{-1}
\\\displaystyle
 {k_e k_{\rm on}\over k_b^{\rm TF}+k_{\rm on}} &
\mbox{for}\quad   (k_b^{\rm TF}+k_{\rm on})^{-1} \lsim t \lsim
t_{\rm pl}
\\\displaystyle
 {k_e k_{\rm on}\over k_b^{\rm TF}+k_{\rm on}} \exp\left[
-{k_b^{\rm TF}k_e\over k_b^{\rm TF}+k_{\rm on}}t
\right] &
\mbox{for }\quad t\gsim  t_{\rm pl}
\end{array}\right.
\end{eqnarray*}
with the plateau time
\begin{equation}
t_{\rm pl}\equiv {1\over k_e}\left( 1+ n_{\rm bst}\right)
\label{S-t_pl}
\end{equation}
and the plateau value
\begin{equation}
C_{\rm max} \approx {k_{\rm on}\over k_b^{\rm TF}+k_{\rm on}}\cdot k_e
= {n_{\rm bst}\over t_{\rm pl}} .
\end{equation}

\subsection{Three step model}

For the three step model, additional complication is that there are
two reversible transitions, i.e. the transition between the off-state
and the closed state, and the transition between the off-state and TF
binding state, thus there exist two sequences of transitions within
each elongation interval.  This can be nicely represented by a
binominal expansion;
\[
\begin{array}{cccc}
 \mbox{state sequence} & \mbox{interval distribution with probability} 
\\ \hline
\mbox{off $\odot$ cl $\bullet$ op - elong, } & 
      \tilde p_{\rm off}(s)q_1 \tilde p_{\rm cl}(s)q_2 \tilde p_{\rm op}(s) \\
 \mbox{$\bigl($ (off$\otimes$TF-)+(off$\odot$cl $\circ$) $\bigr)$ off
 $\odot$ cl $\bullet$ op - elong, }  &
\bigl( \tilde p_{\rm off}(s) p_1 \tilde p_{\rm TF}(s) +
       \tilde p_{\rm off}(s) q_1 \tilde p_{\rm cl}(s) p_2 \bigr)
      \tilde p_{\rm off}(s)q_1 \tilde p_{\rm cl}(s)q_2 \tilde p_{\rm op}(s) \\
 \mbox{$\bigl($ (off$\otimes$TF-)+(off$\odot$cl $\circ$) $\bigr)^2$ off
 $\odot$ cl $\bullet$ op - elong, } &
\bigl( \tilde p_{\rm off}(s) p_1 \tilde p_{\rm TF}(s) +
       \tilde p_{\rm off}(s) q_1 \tilde p_{\rm cl}(s) p_2 \bigr)^2
      \tilde p_{\rm off}(s)q_1 \tilde p_{\rm cl}(s)q_2 \tilde p_{\rm op}(s) \\
 \cdot\cdot\cdot \\
 \mbox{$\bigl($ (off$\otimes$TF-)+(off$\odot$cl $\circ$) $\bigr)^n$ off
 $\odot$ cl $\bullet$ op - elong, } &
\bigl( \tilde p_{\rm off}(s) p_1 \tilde p_{\rm TF}(s) +
       \tilde p_{\rm off}(s) q_1 \tilde p_{\rm cl}(s) p_2 \bigr)^n
      \tilde p_{\rm off}(s)q_1 \tilde p_{\rm cl}(s)q_2 \tilde p_{\rm op}(s) \\
 \cdot\cdot\cdot 
\end{array}
\]
with the period length distributions
\[
\tilde p_{\rm off}(s)={k_b^{\rm TF}+k_b \over s+k_b^{\rm TF}+k_b},\quad
\tilde p_{\rm TF}(s)={k_u^{\rm TF} \over s+k_u^{\rm TF}},\quad
\tilde p_{\rm cl}(s)={k_u+k_o \over s+k_u+k_o},\quad
\tilde p_{\rm op}(s)={k_e \over s+k_e} ,
\]
and the branching probabilities
\[
p_1 = {k_b^{\rm TF}\over k_b^{\rm TF}+ k_b},\quad q_1=1-p_1,\qquad
p_2 = {k_u \over k_u+ k_o},\quad q_2=1-p_2 ,
\]
which are represented by the marks: $\otimes$ for $p_1$, $\odot$ for
$q_1$, $\circ$ for $p_2$, and $\bullet$ for $q_2$.
This gives
\begin{equation}
\tilde p(s) =
{   \tilde p_{\rm off}(s)q_1 \tilde p_{\rm cl}(s)q_2 \tilde p_{\rm op}(s)
\over
1-
\bigl( \tilde p_{\rm off}(s) p_1 \tilde p_{\rm TF}(s) +
       \tilde p_{\rm off}(s) q_1 \tilde p_{\rm cl}(s) p_2 \bigr)
},
\label{S-p3_TF_s}
\end{equation}
and the expression for $p(\tau)$ is
\begin{equation}
p(\tau)  = 
k_b k_o k_e \left[
{k_u^{\rm TF}-k_e \over (k_1-k_e)(k_2-k_e)(k_3-k_e)} e^{-k_e\tau}
+ \sum_{i=1}^3
{k_u^{\rm TF}-k_i \over (k_{i+1}-k_i)(k_{i-1}-k_i)(k_e-k_i)} e^{-k_i\tau}
\right]
\end{equation}
with $-k_i$ ($i=1,2,3$) being the solution of the cubic equation
\begin{equation}
 s^3 + A s^2 + B s + C=0
\end{equation}
with the coefficients
\begin{eqnarray*}
A & \equiv &  k_b^{\rm TF} + k_b + k_u^{\rm TF} + k_u + k_o
\\
B & \equiv & k_u^{\rm TF}(k_u+k_b+k_o) + k_b^{\rm TF}(k_u+k_o) + k_b k_o
\\
C & \equiv & k_b k_o k_u^{\rm TF}.
\end{eqnarray*}
Note that we define $k_i$ as a decay rate with a positive real part.

From eqs.(\ref{S-C-s}) and (\ref{S-p3_TF_s}),
the correlation function $C(t)$ is given by
\begin{equation}
C(t)  = 
{k_b k_o k_e k_u^{\rm TF}\over k^C_1 k^C_2 k^C_3}-
k_b k_o k_e \sum_{i=1}^3
   {(k_u^{\rm TF}-k^C_i)  e^{-k^C_i t} \over 
          k^C_i(k^C_{i+1}-k^C_i)(k^C_{i-1}-k^C_i)} ,
\label{S-3step-C-TF}
\end{equation}
with $-k^C_i$ ($i=1,2,3$) being the solution of the cubic equation
\begin{equation}
 s^3 + D s^2 + E s + F=0
\end{equation}
with the coefficients
\begin{eqnarray*}
D & \equiv & k_b^{\rm TF}+k_b+k_u^{\rm TF}+k_u+k_o+k_e
\\
E & \equiv & 
(k_b^{\rm TF}+k_u^{\rm TF})(k_u+k_o+k_e)+k_b(k_u^{\rm TF}+k_o+k_e)
+(k_u+k_o)k_e
\\
F & \equiv &
k_u^{\rm TF}k_b(k_o+k_e) + (k_b^{\rm TF}+k_u^{\rm TF})(k_u+k_o)k_e .
\end{eqnarray*}

The steady state activity $\Omega_{\rm TF}$ is
\begin{eqnarray}
\Omega_{\rm TF} 
& = &
\lim_{t\to\infty} C(t) 
 = 
{k_b k_o k_e k_u^{\rm TF}\over 
k_u^{\rm TF}k_b(k_o+k_e)+(k_b^{\rm TF}+k_u^{\rm TF})(k_u+k_o)k_e}
\\
& = &
{n_{\rm bst}\over \tau_{\rm bst}+\tau_{\rm TF}}
\quad =
\left\{ \begin{array}{ll}\displaystyle
\Omega_0
\quad
& \mbox{for }{k_b^{\rm TF}\over k_u^{\rm TF}}\to 0
\\ \\ \displaystyle
{n_{\rm bst}\over \tau_{\rm TF}}
%
& \mbox{for }{k_b^{\rm TF}\over k_u^{\rm TF}}\to\infty 
\end{array}\right. ,
\end{eqnarray}
with
\begin{equation}
n_{\rm bst}\equiv {k_b\over k_b^{\rm TF}}\,{k_o\over k_o+k_u},
\qquad
\tau_{\rm bst}\equiv {1\over k_b^{\rm TF}}
           + n_{\rm bst}\left( {1\over k_o}+{1\over k_e}\right),
\qquad
\tau_{\rm TF} = {1\over k_u^{\rm TF}},
\end{equation}
and $\Omega_0$ being the bare activity of the three step model (\ref{S-K3_0}). 
The number of transcriptions $n_{\rm bst}$ in a burst is now given by
the winning ratio of RNAP $k_b/k_b^{\rm TF}$ multiplied by the branching
ratio in the closed state $k_o/(k_o+k_u)$.

The expression for $\Omega_{\rm TF}$ can also be put in the form
analogous to eq.(\ref{S-K3_0}),
\begin{equation}
{1\over \Omega_{\rm TF}}  = 
\left[
{1\over k_b} + {k_b^{\rm TF}\over k_b}\cdot{1\over k_u^{\rm TF}}
\right]
+
\left[
{1\over k_o}+{k_u\over k_o}\cdot\left( 
     {1\over k_b} +{k_b^{\rm TF}\over k_b}\cdot{1\over k_u^{\rm TF}} \right)
\right]
+
{1\over k_e} 
\quad  = 
\tau_b({\rm TF})+\tau_o^*({\rm TF})+\tau_e
\label{S-3step-K_TF}
\end{equation}
with
\begin{equation}
\qquad
\tau_b({\rm TF}) \equiv \tau_b +  {k_b^{\rm TF}\over k_b}\cdot \tau_{\rm TF},
\quad
\tau_o^*({\rm TF}) \equiv
\tau_o + {k_u\over k_o}\cdot \tau_b({\rm TF}),
\qquad
\tau_b={1\over k_b},
\quad
\tau_o = {1\over k_o} .
\end{equation}
This allows a similar interpretation with that for
eq. (\ref{S-2step-K_TF-2}); The average interval of elongation
$1/\Omega_{\rm TF}$ is the sum of three times: (1) $\tau_b(TF)$, the
time for RNAP to bind the promoter, (2) $\tau_o^*(TF)$, the time to form
a open complex for the first time after RNAP binding, and (3) $\tau_e$,
the time to elongate after forming the open complex.  The time
$\tau_b(TF)$ is the sum of (i) $\tau_b$, the RNAP binding time, and (ii)
the unbinding time, $\tau_{\rm TF}$, multiplied by the number of TF
bindings, $k_b^{\rm TF}/k_b$, before an RNAP binds.  Similarly, the time
$\tau_o^*(TF)$ is the sum of (i) $\tau_o$, the time to form an open
complex, and (ii) the binding time $\tau_b(TF)$ multiplied by the
average number of RNAP unbindings, $k_u/k_o$, before it forms an open
complex.

\section{Promoter Interference}

If there is another promoter, pA, competing with the promoter pS in the
parallel or converging position, then their activities interfere with
each other.  Here, we consider only the interference effect on pS by pA.
The activity of pA is given by the elongation interval distribution
$p_A(\tau)$.

We consider two effects for the transcription interference: occlusion
and sitting duck interference.  The occlusion is the effect that an RNAP
cannot bind to the promoter site of pS while the RNAP from pA is passing
over the promoter site.  The time that RNAP needs to pass through the
promoter site is the occlusion time $\tau_{\rm occ}$.  The sitting duck
interference is that the RNAP sitting on the promoter site pS is removed
by the RNAP from pA comes to pS.

Under the interference of pA, the pS activity is limited within the
elongation intervals from pA. Therefore, the average activity of pS
under these effects $\Omega_{\rm TI}$ is the average over the activity within
the interval $\tau$ and the average by the probability that a time is in
the interval of the length $\tau$; This is given by
\begin{equation}
\Omega_{\rm TI}={\displaystyle
\int_{\tau_{\rm occ}}^\infty \!\! p_A(\tau)\,\tau\,
\left(
{1\over \tau}\int_0^{\tau-\tau_{\rm occ}}\!\!\!\!\!\!  C(t)\,\,dt
\right)d\tau
\over\displaystyle
\int_0^\infty p_A(\tau)\,\tau\,  d\tau
} , 
\label{S-K_TI}
\end{equation}
using the transcription initiation correlation function $C(t)$ without
interference effects.

\subsection{Interference with unregulated promoters}

First, we consider the cases where the promoter pS is not regulated by
TF.  In this case, the correlation function $C(t)$ is rather simple
and the interference effect can be represented by a simple factor
$\chi$, that is the averaged fraction of time that is not occluded:
\begin{equation}
\chi \equiv
{\displaystyle
\int_{\tau_{\rm occ}}^\infty
p_A(\tau)\,\tau
          \left( {\tau-\tau_{\rm occ}\over \tau}\right)\,  d\tau 
\over\displaystyle
\int_0^\infty p_A(\tau)\,\tau\, d\tau
}.
\label{S-chi}
\end{equation}
If we assume the simple Poissonian for pA with the activity $\Omega_A$,
\begin{equation}
p_A(\tau) = \Omega_A e^{-\Omega_A\tau},
\label{S-p_A}
\end{equation}
then $\chi$ is given by
\begin{equation}
\chi_P = e^{-\Omega_A\tau_{\rm occ}}.
\label{S-chi_Poisson}
\end{equation}

(i) In the case that pS can be described by the single step model, $C(t)$
is given by the constant $k_e=\Omega_0$ as has been calculated (\ref{S-1step-C}),
thus eq.(\ref{S-K_TI}) gives
\begin{equation}
\Omega_{\rm TI} = 
{\displaystyle
\int_{\tau_{\rm occ}}^\infty \!\! p_A(\tau)\,\tau\,
\left(
{1\over \tau}\int_0^{\tau-\tau_{\rm occ}}\!\!\!\!\!\!  \Omega_0 \,\,dt
\right)d\tau
\over\displaystyle
\int_0^\infty p_A(\tau)\,\tau\,  d\tau
}
\quad =
\chi \Omega_0 ,
\label{S-K_TI-single-step}
\end{equation}
which means that the interference effects reduce the activity by the
factor $\chi$.

(ii) In the case of the two step model with the Poissonian pA,
eq.(\ref{S-K_TI}) can be estimated as
\begin{equation}
\Omega_{\rm TI} =  \chi_P \Omega_0\, {k_{\rm on}+k_e\over k_{\rm on} + k_e + \Omega_A}
\label{S-K_TI-two-step}
\end{equation}
using eq.(\ref{S-2step-C}).
Here, $\Omega_0$ is the bare activity (\ref{S-Omega_0-two-step}) for the two step
model%
\footnote{
Eq.(\ref{S-K_TI-two-step}) disagrees with the corresponding expression 
$$
\Omega_{\rm TI} = 
\Omega_0\,  {\chi (k_{\rm on}+k_e) \over \chi k_{\rm on} + k_e + \Omega_A}
$$
of eq.(4) and Figure 1(d) in Sneppen {\it et al.} [{\it J. Mol. Biol.}
{\bf 346} (2005) 399--409](the notations have been changed from the
original ones).  It is not difficult, however, to see that this
expression cannot be correct because this does not reduce to that for
the single step case (\ref{S-K_TI-single-step}) in the $k_{\rm
on}\to\infty$ limit.  In its derivation, only the on-rate was reduced
by the factor $\chi$, and it was not taken into account that the total
probability of the states available to RNAP is limited by the factor
$\chi$. }.

\subsection{Interference with regulated promoters}

Only difference from the cases above is that we use the correlation $C(t)$
under the effect of TF.
We give explicit expressions only for the Poissonian pA (\ref{S-p_A}).

For the single step promoter pS, using eqs.(\ref{S-1step-K_TF}) and
(\ref{S-K_TI}), we obtain
\begin{equation}
\Omega_{\rm TI} = 
\chi_P \,  \Omega_0 \,\left[
{k_u^{\rm TF}\over k_u^{\rm TF}+ k_b^{\rm TF}} +
   {k_b^{\rm TF}\over k_u^{\rm TF}+ k_b^{\rm TF}} \cdot
      {\Omega_A \over \Omega_A + k_u^{\rm TF}+ k_b^{\rm TF} }
\right]
\quad
=
\chi_P \, {n_{\rm bst}\over \displaystyle
  \tau_{\rm bst}+{\tau_{\rm TF}\Omega_A^{-1}\over \tau_{\rm TF}+\Omega_A^{-1}}
}
\end{equation}
with $\chi_P$ given by eq.(\ref{S-chi_Poisson})
and $n_{\rm bst}$ and $\tau_{\rm bst}$ by eq.(\ref{S-n_bst-1}).
The last expression simply shows that the quiescent period is interrupted
to be $\Omega_A^{-1}$ when $\Omega_A^{-1}< \tau_{\rm TF}$.

For the two step promoter, we give the expression only for $\tau_{\rm
occ}=0$, namely, without the occlusion effect:
\begin{eqnarray*}
\Omega_{\rm TI} & =  &
{k_e k_{\rm on} k_u^{\rm TF} \over k_+^Ck_-^C} +
 {k_e k_{\rm on} \over k_+^C - k_-^C}
\left[
{k_-^C - k_u^{\rm TF}\over k_-^C} {\Omega_A \over \Omega_A + k_-^C}
- 
{k_+^C - k_u^{\rm TF}\over k_+^C} {\Omega_A \over \Omega_A + k_+^C}
\right]
\\
& = &
\Omega_{\rm TF}
\left[
1 -
{\Omega_A ( \Omega_A + k_+^C + k_-^C - k_+^Ck_-^C/k_u^{\rm TF}) \over
 (\Omega_A + k_-^C) (\Omega_A + k_+^C) }
\right]
\label{S-2step-K_TI}
\end{eqnarray*}
with $k_\pm^C$ given by eq.(\ref{S-2step-k_pmC}) and $\Omega_{\rm TF}$ being
the steady activity under TF by eq.(\ref{S-2step-K_TF}).

In the same approximation as eq.(\ref{S-2step-C-TF-app}),
 $k_b^{\rm TF}\gg k_e \gg k_u^{\rm TF}\approx 0$,
using the approximate form of $k_\pm^C$, we obtain
\begin{eqnarray*}
\Omega_{\rm TI} 
& \approx &
 k_{\rm on}\,
   {k_e\over \Omega_A + k_b^{\rm TF}k_e/(k_b^{\rm TF}+k_{\rm on})}\cdot
   {1\over 1 + (k_b^{\rm TF}+k_{\rm on})/\Omega_A}
\quad =
{n_{\rm bst}\over t_{\rm pl}+\Omega_A^{-1}}\cdot
     {k_b^{\rm TF}+k_{\rm on}\over k_b^{\rm TF}+k_{\rm on}+\Omega_A}
\\
& \approx & \left\{ \begin{array}{ll}\displaystyle
n_{\rm bst} \Omega_A &
\mbox{for }
   \Omega_A\ll 1/t_{\rm pl}={k_b^{\rm TF}\over k_b^{\rm TF}+k_{\rm on}}k_e
\\\displaystyle
{n_{\rm bst}\over t_{\rm pl}}
&
\mbox{for }
1/t_{\rm pl} \ll \Omega_A \ll ( k_b^{\rm TF}+k_{\rm on} )
\\\displaystyle
{n_{\rm bst}\over t_{\rm pl}}\,{k_b^{\rm TF}+k_{\rm on}\over \Omega_A}=
{k_{\rm on}k_e\over \Omega_A} &
\mbox{for }
  ( k_b^{\rm TF}+k_{\rm on} ) \ll \Omega_A
\end{array}\right.  
\end{eqnarray*}
with $n_{\rm bst}$ (\ref{S-n_bst-2}) and $t_{\rm pl}$ (\ref{S-t_pl}).
The maximum value of $\Omega_{\rm TI}$ is achieved at
$\Omega_A\approx\sqrt{k_b^{\rm TF}k_e}$.  The behavior of $C(t)$ in
(\ref{S-2step-C-TF-app}) and $\Omega_{\rm TI}$ as a function of $\Omega_A$
correspond to each other by the correspondence $t\sim 1/\Omega_A$.

For the three step model, we give only a formal solution for
Poissonian pA:
\begin{equation}
\Omega_{\rm TI}   = 
{k_b k_o k_e k_u^{\rm TF}\over k^C_1 k^C_2 k^C_3}-
k_b k_o k_e \sum_{i=1}^3
{(k_u^{\rm TF}-k^C_i)  \over k^C_i(k^C_{i+1}-k^C_i)(k^C_{i-1}-k^C_i)}\, 
{\Omega_A\over \Omega_A+k^C_i} ,
\end{equation}
with the same $k_i^C$ ($i=1,2,3$) with those in eq.(\ref{S-3step-C-TF}).
\renewcommand{\theequation}{A.\arabic{equation}}
\setcounter{equation}{0}

\renewcommand{\thesection}{Appendix:}

\section{Mathematical explanation for $\tau_o^*$ in (\ref{S-K3_0})}

In this appendix, we will give a mathematical explanation for the expression
of $\tau_o^*$ in eq.(\ref{S-K3_0}).

This is the time for RNAP and promoter to form the open complex after
the first binding of RNAP.  Since the initial binding/unbinding
process is reversible, after the first binding, the system may either go
forward to form the open complex, or may go backward to unbind with
the probabilities,
\begin{equation}
q={k_o\over k_o + k_u}
\quad\mbox{and}\quad
p={k_u\over k_o + k_u},
\end{equation}
respectively.  The average waiting time for either of the cases to happen is
\begin{equation}
{1\over k_o + k_u}.
\end{equation}
If the system goes backward to unbind the RNAP, then after the time
\begin{equation}
{1\over k_b},
\end{equation}
another RNAP binds to form the closed complex again, and the situation
becomes the same as before.

If the system goes backward to unbind $n$ times before it proceeds to
form the open complex, the times that the system spends and the
probabilities that should occur are given
\begin{equation}
\begin{array}{ccc}\displaystyle
n & \mbox{time} & \mbox{probability} \\
\hline
1 &\displaystyle {1\over k_o + k_u} & q \\
2 &\displaystyle {1\over k_o + k_u}+{1\over k_b} + {1\over k_o + k_u} & p\, q \\
& \cdot\cdot\cdot  & \\
n &\quad\displaystyle \left({1\over k_o + k_u}+{1\over k_b}\right)n + {1\over k_o + k_u} & p^n\, q \\
& \cdot\cdot\cdot  & 
\end{array}
\end{equation}
thus the average time is given by
\begin{equation}
\sum_{n=0}^\infty
\left[
 \left({1\over k_o + k_u}+{1\over k_b}\right)n + {1\over k_o + k_u} 
\right] p^n\, q 
=
{1\over k_o} + {k_u\over k_o}\, {1\over k_b}
\end{equation}
which is $\tau_o^*$ in eq.(\ref{S-K3_0}).


\end{document}